\documentclass[aps, prb, groupedaddress, preprint, showkeys]{revtex4-1}

\usepackage{graphicx}
\usepackage[]{natbib}
\usepackage{hyperref}
\usepackage{todonotes}
\usepackage{bm}
\usepackage{chngcntr}

\makeatletter
\renewcommand{\@seccntformat}[1]{}
\makeatother

\newcommand{\ie}{i.e.~}

\newcommand{\etal}{et al. ~}
\newcommand{\cf}{cf. ~}
\newcommand{\eg}{e.g.~}
\graphicspath{{./Figures/}{../Figures/}}
\newcommand{\hc}{{\rm h. \, c.}}
\newcommand{\Rsk}{R_{\rm sk}}
\newcommand{\mjs}{m_{\rm J}^*}
\newcommand{\mus}{\mu^*}

\newcommand\Eq[1]{Eq.~(\ref{#1})}
\newcommand\Fig[1]{Fig.~\ref{#1}}

\usepackage{amsmath}
\usepackage{amsfonts}
\usepackage{amssymb}
\usepackage{amsthm}
\usepackage{lipsum}
\renewcommand{\H}{{\cal H}}
\newcommand{\up}{\uparrow}
\newcommand{\down}{\downarrow}
\renewcommand{\hc}{{\rm h. \, c.}}

\begin{document}

\title{Topological superconductivity with deformable magnetic skyrmions}

\author{Maxime Garnier}
\email[Corresponding author: \,]{maxime.garnier1@u-psud.fr}
\affiliation{Laboratoire de Physique des Solides, UMR 8502, CNRS,
Universit\'{e} Paris-Sud, Universit\'{e} Paris-Saclay, 91405 Orsay, France}
\author{Andrej Mesaros}
\affiliation{Laboratoire de Physique des Solides, UMR 8502, CNRS,
Universit\'{e} Paris-Sud, Universit\'{e} Paris-Saclay, 91405 Orsay, France}
\author{Pascal Simon}
\affiliation{Laboratoire de Physique des Solides, UMR 8502, CNRS,
Universit\'{e} Paris-Sud, Universit\'{e} Paris-Saclay, 91405 Orsay, France}

\date{\today}

\begin{abstract}
Magnetic skyrmions are nanoscale spin configurations that are efficiently created and manipulated. They hold great promises for next-generation spintronics applications.
In parallel, the interplay of magnetism, superconductivity and spin-orbit coupling has proved to be a versatile platform for engineering topological superconductivity predicted to host non-abelian excitations, Majorana zero modes. 
We show that topological superconductivity can be induced by proximitizing skyrmions and conventional superconductors, without need for additional ingredients.
Apart from a previously reported Majorana zero mode in the core of the skyrmion, we find a more universal chiral band of Majorana modes on the edge of the skyrmion. We show that the chiral Majorana band is effectively flat in the physically relevant parameter regime, leading to interesting robustness and scaling properties. In particular, the number of Majorana modes in the (nearly-)flat band scales with the perimeter length of the system, while being robust to local disorder.
\end{abstract}

\maketitle

\section{Introduction}

Magnetic skyrmions are nano- or meso-scale whirling spin configurations of topological nature which gives them some stability and long lifetime. Magnetic skyrmions have been found in a variety of non-centrosymmetric magnets \cite{TokNag2013}, 
in ultrathin magnetic films \cite{Heinze2011,Romming2013,Gross2018}  as well as in multiferroic insulators \cite{Seki2012,Tokura2012,White2012}. Quite remarkably, magnetic skyrmions can be stabilized over a wide temperature domain ranging from room temperature 
\cite{Yu2011,Parkin2017} to cryogenic temperature \cite{Heinze2011,Romming2013,Romming2015}.
Evidence that magnetic skyrmions can be driven by ultralow electric current densities \cite{Jonietz2010,Yu2012} make them 
promising candidates for future spintronic applications
\cite{Fert2013}.

In parallel to these developments, the search for Majorana modes in condensed-matter systems has been the focus of great attention, motivated by their potential application in quantum computation. Various systems have been considered as hosts for topological superconductivity and Majorana modes, based on the paradigm of combining ferromagnetic order with strong spin-orbit coupling and conventional superconductivity. The paradigm led to successes in predicting \cite{Nakosai2013,NP2013,Braunecker2013,Klinovaja2013,Pientka2013} and experimentally indicating Majorana zero-energy modes at endpoints of one-dimensional systems, such as iron atomic chains \cite{Nadj-Perge_Yazdani_adatom_chain_2014,Pawlak2016,Wiesendanger_Chain_2018} and semiconducting wires\cite{Lutchyn:2018}. Recent experiments have extended the paradigm to two dimensions, reporting some evidence for dispersive Majorana edge states around two-dimensional magnetic domains using cobalt atom clusters under monolayer lead\cite{MenardPbCoSi} or iron adatom clusters on a rhenium surface\cite{PalacioMoralesWiesendanger_Island_2018}. Since the long-term goal is a flexible platform for manipulation of Majorana modes, two challenges for the paradigm are that the preformed structures (clusters, wires) are hard to manipulate, and that the systems are constrained by requirement of strong spin-orbit coupling.

An alternative approach to engineering topological superconductors while circumventing these two challenges could be to remove the spin-orbit coupling ingredient, and instead consider a non-collinear magnetic texture proximitized by a conventional superconductor\cite{Nakosai2013,Schnyder2015,Loss_Majorana_skyrmion, Zutic_PRL, Mohanta2019, Kontos2019}. In fact, our results are relevant for a broader class of skyrmion-like textures, such as magnetic bubbles\cite{Kiselev:2011, Tchernyshyov:2012, Buttner:2018, BernandMantel:2018}. Additionally, a texture such as a skyrmion can be manipulated by external fields, potentially facilitating the manipulation of Majorana states. Yang \etal recently found that skyrmions having an even azimuthal number can indeed bind a single Majorana zero mode in their core \cite{Loss_Majorana_skyrmion}. Moreover, very elongated magnetic skyrmions were shown to host Majorana zero modes at their endpoints\cite{KovalevPRB2018}.
  In contrast, we find here that a magnetic skyrmion of any azimuthal winding and sufficient radial winding gives rise to a single band of states at the edge of the skyrmion, \ie, a chiral Majorana edge mode (CMEM). Surprisingly, for the physically relevant range of parameters (skyrmion size, winding numbers, magnetic coupling strength) the CMEM has negligible velocity, \ie, it is nearly a Majorana flat band (MFB). Furthermore, we find that the CMEM is robust to local perturbations, as well as to smooth deformations of the edge geometry. Such deformations preserve the number of edge states proportional to the perimeter length of the edge.

For systems with translational symmetry there is a theoretical classification of topological superconducting phases, and predictions for a corresponding CMEM along a given edge of the system\cite{WongLee12,HaoTing16,DaidoYanase17}. Furthermore, the existence of a MFB along an edge can be deduced from an appropriate discrete chiral symmetry and topological indices in lower spatial dimension\cite{Volovik11,Sato11,WangLee12}. In our case, the skyrmion is an inhomogeneous texture so these methods cannot be directly used to explain the observed robustness and near-flatness of the CMEM. We however deduce the underlying topological protection of the skyrmion's CMEM by a mapping to a cylinder geometry. Although this construction requires rotational symmetry of the skyrmion, the CMEM by its nature provides robustness against small deformations of the shape of the system. Further, we identify the chiral symmetry that would protect a strict MFB (instead of a CMEM), and show that this symmetry is only weakly broken by the skyrmion texture, leading to a nearly flat CMEM and providing further protection against low-energy perturbations. Finally, we will discuss potential material realizations, and possibilities for manipulation of Majorana states within the nearly-flat CMEM.

\section{Results}
\subsection{Setup and model}

\begin{figure}[h!]
\centering
\hspace*{-1.cm}
\includegraphics[scale=.95]{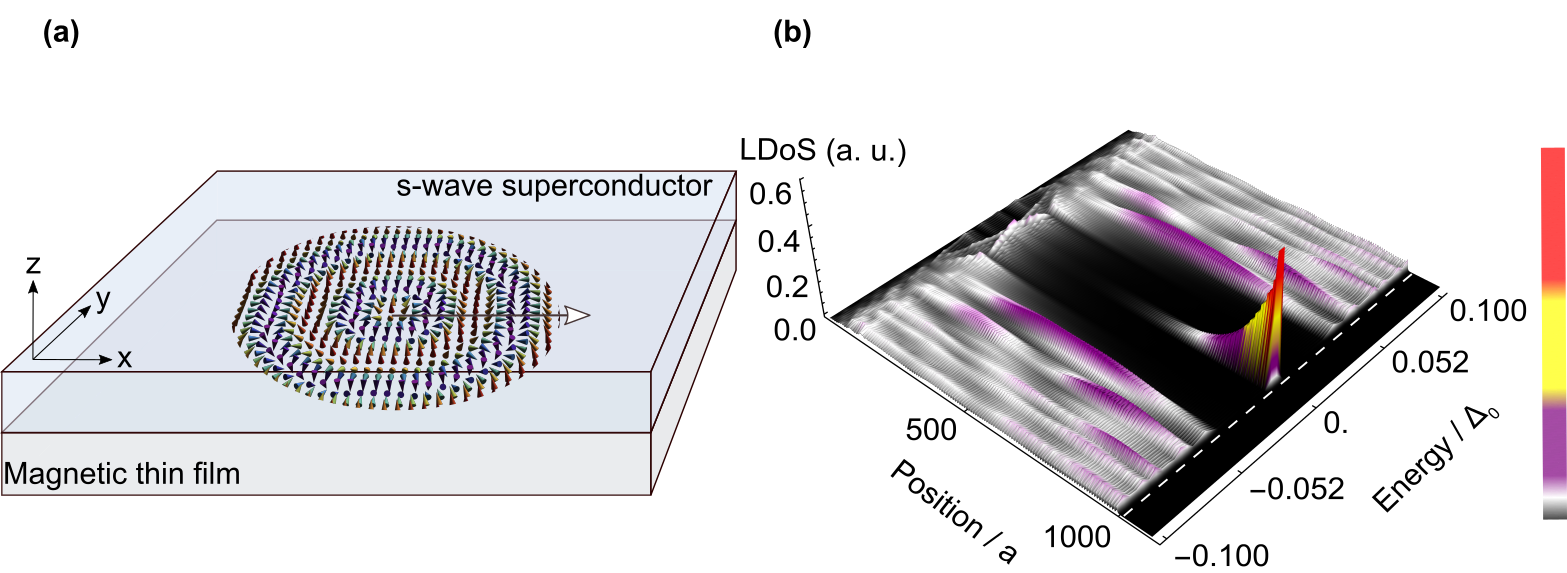}
\caption{{\bf Induced topological superconductivity} {\bf a} A 2D magnetic thin film hosting a skyrmion (bottom layer) with radial winding number $p = 4$ and azimuthal winding number $q = 1$ proximitized by a $s$-wave superconductor (top layer).
Arrow colors represent the $z$ component of the skyrmion texture (red for up, purple for down).
{\bf b} Local density of states (LDoS) along the gray line in {\bf a} obtained from a
tight-binding description with lattice spacing $a=1$ and hopping energy $t=1/(2m a^2)$ with electron effective mass $m$ (see Methods Radial tight-binding model). The LDoS shows very weakly dispersing edge states within an effective gap $\Delta_{\rm eff} \approx 5 \% \Delta_0$. The colorbar refers to the LDoS going from low values (black) to high values (red). Beyond the edge of the skyrmion (vertical white dashed line) we include a non-magnetic background. The model parameters are the $s$-wave order parameter $\Delta_0/t=0.1$, the exchange interaction strength $J/\Delta_0=2$ and the chemical potential $\mu/t = 0$, with an angular momentum cutoff set to 50, see Methods Radial tight-binding model. The skyrmion has radial winding number $p=6$, azimuthal winding number $q=2$, and radius $R_{\rm sk}/a = 996$, so that the length of a spin flip $\lambda = R_{\rm sk}/p$ is $\lambda/a =  166$.}
\label{fig: fig1}
\end{figure}

Consider a two-dimensional (2D) magnetic thin film hosting a skyrmion, which is represented by a classical magnetization texture
\begin{align}
 {\bf n}\left({\bf r}\right)  = \left(\sin f(r) \cos ( q\theta), \sin f(r) \sin ( q\theta ), \, \cos f(r)\right),
\label{eq: skyrmion_def}
\end{align}
written in polar coordinates ${\bf r} = (r, \theta)$, where $f(r)$ is a radial profile that we will specify shortly. We study such a thin film proximitized by a conventional $s$-wave superconductor (\Fig{fig: fig1}a).
The electrons interact with the skyrmion texture via a direct exchange interaction of strength $J$. In the Nambu basis $\Psi^\dagger \left({\bf r}\right) = \left(\psi_\up^\dagger\left({\bf r}\right) , \psi_\down^\dagger\left({\bf r}\right),   \psi_\down\left({\bf r}\right) , -\psi_\up\left({\bf r}\right) \right)$, where $\psi_\sigma\left({\bf r}\right) $ annihilates an electron with spin $\sigma$ at position ${\bf r}$ in 2D, the total Hamiltonian $H$ can be written as $H = \frac{1}{2}\int d{\bf r} \, \Psi^\dagger \left({\bf r}\right)  \H\left({\bf r}\right)  \Psi \left({\bf r}\right) $, with the Bogoliubov-de-Gennes (BdG) Hamiltonian $\H\left({\bf r}\right) $ defined as
\begin{align}
\H\left({\bf r}\right)  = \left(-\dfrac{\nabla^2}{2m} - \mu\right) \tau_z + J	\, \bm{\sigma} \cdot {\bf n}\left({\bf r}\right)  + \Delta_0 \, \tau_x,
\label{eq: BdG_Skyrmion}
\end{align}
where the $s$-wave superconducting order parameter $\Delta_0$ is taken real without loss of generality, the electron effective mass is $m$ and the chemical potential is $\mu$. We set $\hbar = 1$ unless explicitly written otherwise. The $\sigma_\alpha$ and $\tau_\alpha$ ($\alpha = x,\, y,\, z$) are Pauli matrices acting in spin and particle-hole space, respectively. 

We thus assume that the skyrmion affects the electrons only through the exchange field, which is justified in the limit of strong local exchange interaction. We further consider the limit where the effective spin-orbit coupling is dominated by the one induced by the skyrmion exchange field. This is justified if both intrinsic and Rashba spin-orbit couplings are relatively weak, which we argue in the discussion section to be the case in a typical superconducting material such as aluminium. The inclusion of intrinsic and Rashba spin-orbit couplings would require another in-depth study due to the loss of rotational symmetry, and due to the non-trivial interplay of different effective spin-triplet pairings introduced by these couplings. In the limit that we consider in this work, in absence of skyrmion the spin-orbit length in the superconductor $l_{\rm so} = 1/(m\alpha)$, with $\alpha$ a spin-orbit amplitude, is much larger than the typical lengthscale of the skyrmion, so that the magnetoelectric coupling and appearance of vortices can also be neglected \cite{Rudner2016, Buzdin2018,Takashima2016, Eremin2018, Rex_SK_Vortex_2019}.

The skyrmion is parametrized by three numbers: the radial winding number $p$, which counts the number of spin flips as one moves radially away from the core of the skyrmion;  the azimuthal winding number $q$, which counts the number of spin flips as one winds around the origin; and finally the skyrmion radius $R_{\rm sk}$, which determines its size. We consider a hard-wall boundary condition at the edge \ie at $r=R_{\rm sk}$. Formally this can be realized by having the exchange $J=0$ in the magnetic insulator outside the edge, which might be experimentally unattainable. However a simple alternative is to deposit the superconductor in form of an island, whose edge would naturally become the edge in our model. In such a setup it is natural to consider various geometrical shapes of the edge given a fixed underlying magnetic texture.

For simplicity, the function $f(r)$ is chosen to be linear, defining a straightforward skyrmion texture as in Fig.~\ref{fig: fig1}a. As we show in the discussion section, the exact shape of $f(r)$ has weak influence on our conclusions, allowing us to extend our results to a broader class of textures including magnetic bubbles.

\subsection{Skyrmion edge states and topological superconductivity}
We first solve the model in \Eq{eq: BdG_Skyrmion} by using rotational symmetry. Our model in \Eq{eq: BdG_Skyrmion} has a rotational symmetry in the combined real- and spin-space, given by the conserved total angular momentum $J_z = L_z + \frac{q}{2} \sigma_z$ where $ L_z = -i\partial_\theta$ is the orbital angular momentum. Using the eigenvalues of $J_z$, denoted as $m_{\rm J}$, the Hamiltonian in \Eq{eq: BdG_Skyrmion} becomes an effective one-dimensional radial model that can further be discretized and diagonalized numerically (technical details are given in Methods Radial tight-binding model). In the regime of $J/\Delta_0$ large enough (estimated as $J>\sqrt{\Delta_0^2+\mu^2}$ below), \Fig{fig: fig1}b shows the resulting strong peak in the local density of states (LDoS), near zero energy and at the skyrmion's edge. Such a spectral feature was observed before in related models \cite{Nakosai2013, Schnyder2015,Loss_Majorana_skyrmion, Morr_engineering}. Further, the LDoS clearly displays a reduced gap $\Delta_{\rm eff} \approx 5 \% \, \Delta_0$ consistent with an effective (topological) $p$-wave superconducting gap\citep{AliceaPRB}. As $J/\Delta_0$ is reduced the effective gap closes (at $J=\sqrt{\Delta_0^2+\mu^2}$, see estimate below), and a full gap $\Delta_0$ develops without any edge states. This is expected in a transition from a topological superconducting phase to trivial superconductivity. The regime of topological $p$-wave superconductivity is also consistent with our finding of other in-gap states localized near the skyrmion core that only appear when the edge modes appear. We therefore interpret these states near the core as analogs of states bound to magnetic impurities (here, inhomogeneities of the skyrmion texture), which are only expected for $p$-wave pairing, but are absent in the $s$-wave-pairing-dominated trivial phase ($J/\Delta_0$ small enough).

We further clarify the edge states and topological superconductivity by looking at the spectrum $\varepsilon(m_{\rm J})$, in which edge states form a seemingly flat band in a range of $m_{\rm J}$ values, see \Fig{fig: fig2}a and b. Importantly, the edge states appear for any value of the azimuthal winding number $q$ (on the other hand, $p$ always needs to be high enough\cite{Loss_Majorana_skyrmion}, we showcase $p = 6$). Note that in contrast, we find a single Majorana zero mode at the core of the skyrmion only if the skyrmion's azimuthal winding number $q$ is even\cite{Loss_Majorana_skyrmion}. This is easily understood since the zero mode must appear in the self-conjugate angular momentum $m_{\rm J}=0$ sector, while $m_{\rm J}$ is quantized to be integer (resp. half-odd-integer) when $q$ is even (odd) due to the single-valuedness of the wavefunction. The existence of edge states indicates that skyrmions of any $q$ induce topological superconductivity. 

\subsection{Topological origin and the near-flatness of edge mode}
\begin{figure}[h!]
\centering
\hspace*{-.8cm}
\includegraphics[scale=0.8]{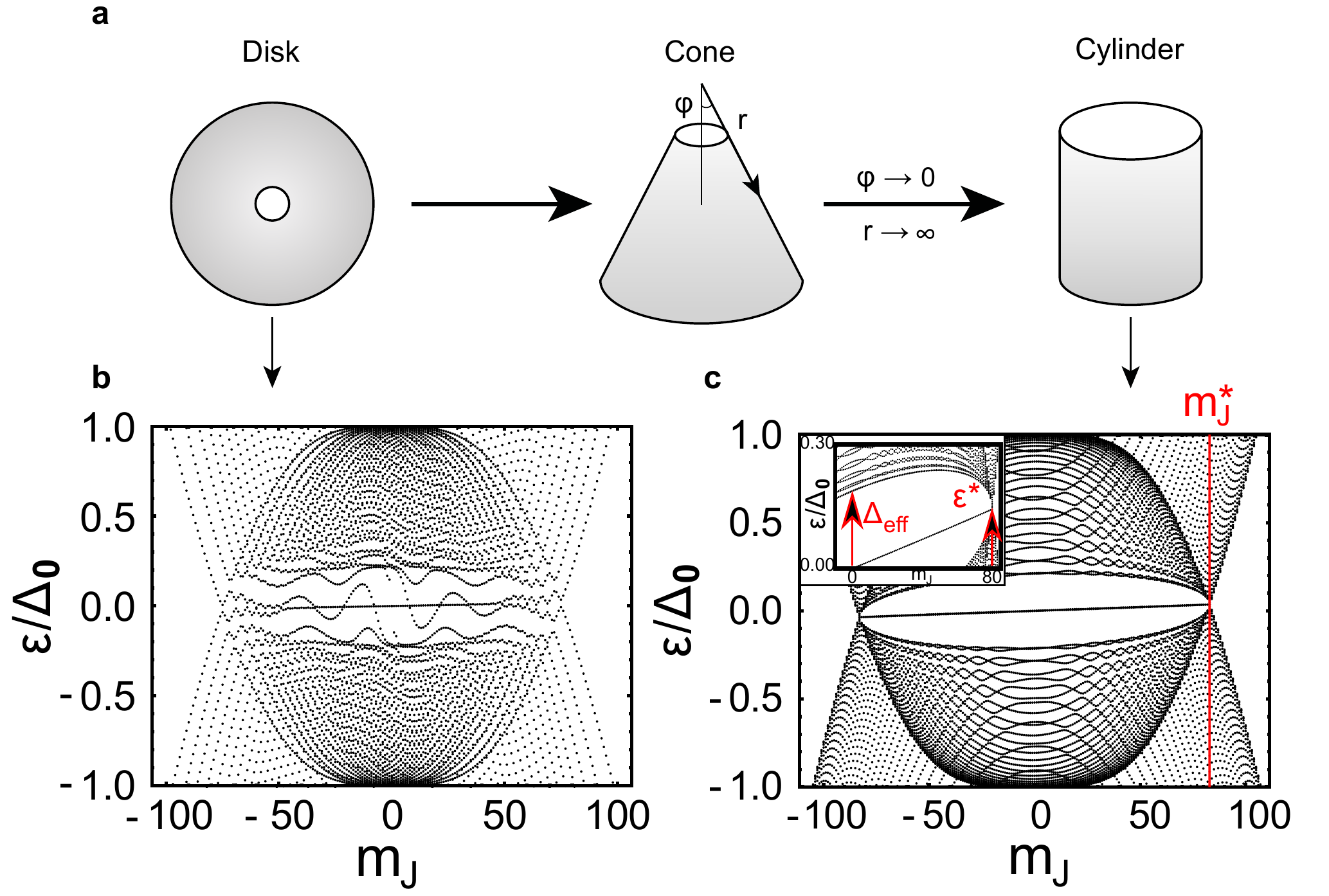}
\caption{{\bf Mapping the skyrmion from a disk to a cylinder.} {\bf a}
The half-opening angle $\varphi$ introduces the mapping: $\varphi = \pi/2$ realizes the disk geometry while the limit $\varphi \to 0$ with $r\to \infty$ and $r \sin\varphi = \Rsk$ realizes the cylinder geometry, where $r$ is the distance with respect to the tip of the cone. The core is covered by a white disc for clarity. The excitation energy spectrum $\varepsilon$ of the original skyrmion model {\bf b} and the model on the cylinder {\bf c}, as function of the angular momentum quantum number $m_{\rm J}$ (see Methods Radial tight-binding model). At $\pm \mjs$ (vertical red lines) the bulk gap closes. (Inset) Zoom-in of the spectrum, defining the effective gap $\Delta_{\rm eff}$ and the maximal energy reached by the edge mode, $\varepsilon^*$. The parameters of the computations are the skyrmion radius $R_{\rm sk}/a = 200$, the chemical potential $\mu/t = 0$, the $s$-wave pairing amplitude $\Delta_0/t=0.1$, the exchange interaction strength $J/t = 0.2$, radial winding number $p = 6$ and azimuthal winding number $q=2$ for {\bf b}. For {\bf c}, the cylinder radius and height are both $200 \, a$ while $\mu/t = 0$, $\Delta_0/t=0.1$, $J/\Delta_0 = 2$, $p = 6$ and $q = 2$.}
\label{fig: fig2}
\end{figure}

In order to explain the origin of the edge states we use a procedure introduced by Wu et al.\cite{MappingMartin} to smoothly deform the model in \Eq{eq: BdG_Skyrmion} defined on the disk to another model defined on the cylinder via the cone geometry as represented in \Fig{fig: fig2}a. 
Taking the cylinder limit (see Methods Gradient and Laplace operators in the cone geometry) effectively focuses on the edge of the skyrmion at the price of disregarding the skyrmion core area, which is replaced by an artificial edge.

Explicitly, we use the rotation symmetry, \ie the total angular momentum $m_{\rm J}$ basis (see Methods Radial tight-binding model), then we apply the unitary transformation $U(r) = \exp(i\sigma_yf(r)/2)$ to align the exchange field with the $z$-axis at each point of radial distance $r$, and finally we apply the mapping to the cylinder. The resulting Hamiltonian $\widetilde{\H}_{m_{\rm J}}^{\rm cyl}(r)$ can be written as the sum of three parts, $\widetilde{\H}_{m_{\rm J}}^{\rm cyl}(r) = \H_{m_{\rm J}}^{\rm wire}(r) + \H_{m_{\rm J}}^{\rm slope}(r) + \H_{m_{\rm J}}'(r)$, where:
\begin{widetext}
\begin{align}
  \label{eq: BdG_Hamiltonian_cont_Cylinder_wire}
  &\H_{m_{\rm J}}^{\rm wire}(r) = \left[- \frac{1}{2m} \partial_r^2 - \mu\right]\tau_z + \frac{1}{2mR_{\rm sk}^2} \left(m_{\rm J}^2 + \frac{q^2}{4}\right)\tau_z + \frac{f'}{2m}  \partial_ri\sigma_y \tau_z + J \sigma_z + \Delta_0 \tau_x\\
\label{eq: BdG_Hamiltonian_cont_Cylinder_slope}
&\H_{m_{\rm J}}^{\rm slope}(r)=-\frac{q\, m_{\rm J}}{2 m R_{\rm sk}^2}(-1)^p \sigma_z\tau_z\\
\label{eq: BdG_Hamiltonian_cont_Cylinder_corr}
&\H_{m_{\rm J}}'\left(r\right) = \frac{f'^2}{8m}\tau_z + \frac{f''}{4m} i\sigma_y \tau_z 
\end{align}
\end{widetext}
For our purpose it is sufficient to show that the edge modes and the effective gap (the energy gap in the $m_{\rm J}=0$ sector) of the original model \Eq{eq: BdG_Skyrmion} are connected to such features of a model derived from the cylinder mapping. Therefore, in what follows we safely neglect the part in \Eq{eq: BdG_Hamiltonian_cont_Cylinder_corr} since these are a small overall chemical potential renormalization and small overall boundary term.
For a given angular momentum $m_{\rm J}$, the $\H_{m_{\rm J}}^{\rm wire}(r)$  can be interpreted as the extensively studied Hamiltonian of a Rashba wire \cite{Lutchyn2010,Oreg2010} upon introducing a momentum-dependent chemical potential $\mu\left(m_{\rm J}\right) =  \mu -  \left(m_{\rm J}^2+\frac{q^2}{4}\right)/\left(2mR_{\rm sk}^2\right)$. (Note however that the skyrmion-induced effective spin-orbit coupling in 2D is not of a simple Rashba type.) At each $m_{\rm J}$ the superconducting wire Hamiltonian $\H_{m_{\rm J}}^{\rm wire}(r)$ is well known to be in a trivial state ($J<\sqrt{\Delta_0^2+\mu(m_{\rm J})^2}$) or in a topological state\cite{Lutchyn2010,Oreg2010} ($J>\sqrt{\Delta_0^2+\mu(m_{\rm J})^2}$). For each topological wire there is a single Majorana zero mode localized at the end of the wire, \ie a single zero mode at the edge of the skyrmion. Due to the variation of $\mu(m_{\rm J})$, there is generically a flat zero-energy band of edge modes, \ie a MFB, for a range of $|m_{\rm J}|<\left\vert m_{\rm J}^*\right\vert$, where
\begin{align}
  \label{eq: critical_m}
\left\vert \mjs\right\vert = R_{\rm sk}\sqrt{\mu + \sqrt{J^2-\Delta_0^2}},
\end{align}
where all energies are in units of the bandwidth $t$, all distances are in units of the lattice spacing $a$ (see Supplementary Note 1 and Supplementary Figure 1 for details). For precisely $|m_{\rm J}|=\left\vert \mjs \right\vert$ the wire is at the topological transition and has a gapless spectrum, giving our model a bulk-gap-closing point as shown in \Fig{fig: fig2}c.

The MFB found here has a protection by a chiral symmetry, as MFB's were found to have in models with translational symmetries\cite{SedlmayerBena11,Sato11,WangLee12}. Note that the wire Hamiltonian and its MFB become a correct model for our texture Eq.~\eqref{eq: skyrmion_def} if we choose $q=0$ and thereby nullify the $\H_{m_{\rm J}}^{\rm slope}(r)$ term. Physically, this is a special case where instead of the skyrmion shape the texture becomes coplanar (in the $xz$-plane, see \Eq{eq: skyrmion_def}), and the orthogonal direction provides a chiral operator
\begin{equation}
  \label{eq: chiral_operator}
  \Xi=\tau_y\sigma_y
\end{equation}
that anticommutes with the Hamiltonian (see \Eq{eq: BdG_Skyrmion}). Since all the MFB states have the same chirality, they cannot hybridize among themselves. It is difficult to remove the MFB states\cite{SedlmayerBena11}, namely, a perturbation must have energy larger than the effective gap; or, it should hybridize the MFB with low energy bulk states at  $|m_{\rm J}|\approx\left\vert \mjs \right\vert$, which are few; or, chirality symmetry must be broken (out-of-$xz$-plane exchange field). We note that the proof of existence of the MFB rests on the rotational symmetry of the $q=0$ coplanar texture, since this symmetry provides the $m_{\rm J}$ quantum number. Consider now deformations of the shape of the edge imposed on our $q=0$ coplanar texture. These geometric deformations would generally mix the $m_{\rm J}$ sectors, yet the described stability of the MFB implies that the deformations would be inefficient in removing the MFB states.

We can now proceed to the relevant model for a skyrmion with arbitrary $q\neq0$:
\begin{equation}
  \label{eq:2}
\widetilde{\H}_{m_{\rm J}}^{\rm cyl,eff}(r)=\H_{m_{\rm J}}^{\rm wire}(r) + \H_{m_{\rm J}}^{\rm slope}(r).
\end{equation}
The single term $\H_{m_{\rm J}}^{\rm slope}(r)$ breaks the chiral symmetry $\Xi$, and there are no other chiral operators. The term $\H_{m_{\rm J}}^{\rm slope}(r)$ exactly contributes an energy $\varepsilon^{\rm edgestate}(m_{\rm J})\sim m_{\rm J}$ to an MFB state at $m_{\rm J}$, making the flat MFB into a linearly dispersing chiral Majorana edge mode (CMEM) of the $q\neq0$ skyrmion (\Fig{fig: fig2}c). The single CMEM itself has general robustness to perturbations, however, we additionally find that the velocity of the CMEM is very small in the relevant physical regime, i.e., the breaking of chiral symmetry is very weak. Qualitatively, we can estimate the upper limit on energy $\varepsilon^*$ that the CMEM can have, which occurs at the maximal $m_{\rm J}$ of the CMEM, \ie, $\varepsilon^*\equiv\vert \varepsilon^{\rm edgestate}(\left\vert \mjs\right\vert)\vert$. Treating $\H_{m_{\rm J}}^{\rm slope}(r)$ as a first order perturbation to the MFB (see Supplementary Note 2 and Supplementary Figure 2), the estimate $\varepsilon^*= \frac{q}{\Rsk}\sqrt{\mu+\sqrt{J^2-\Delta_0^2}}$ scales the same way with skyrmion size as the estimate of the effective gap $\Delta_{\rm eff}\sim p/R_{\rm sk}$. For the relevant regime of $J,\, \Delta_0, \, \mu$ (see Discussion) the quantitative ratio is at most $\varepsilon^*/\Delta_{\rm eff}\sim0.1$. The corresponding Fermi velocity of the CMEM is therefore small and suppressed by the skyrmion size, $\partial \varepsilon^{\rm edgestate}(m_{\rm J})/\partial (m_{\rm J}/R_{\rm sk})\sim\frac{q}{2m R_{\rm sk}}$.

We thus demonstrated that at low energy the single edge mode of the skyrmion can be connected to the single CMEM of a cylinder made of Rashba wires, and the CMEM is nearly a MFB. The phase diagram of both models (skyrmion model vs. wires on cylinder) obtained by varying $J/t$ in the radial tight-binding setup are compared in Supplementary Note 3 and Supplementary Figure 3 and show excellent agreement. Importantly, in both the original skyrmion and the cylinder model and for small enough systems as shown in \Fig{fig: fig2}b, we observe the angular momentum value $\mjs$ in accordance to predictions in \Eq{eq: critical_m}, and we observe the near-flatness of the edge mode.

\subsection{Edge states on deformed edges}
The number of states in the single CMEM of a perfectly rotationally symmetric system is given simply by the highest angular momentum that is reached by the edge states, \ie $\mjs$, and therefore scales linearly with the perimeter of a disk-shaped system centered on the rotationally symmetric skyrmion (see \Eq{eq: critical_m}, neglecting corrections of order 1 due to $\mu$ depending on $\mjs$). We remind that the edge of system is defined by setting exchange $J=0$ outside it, or equivalently, by depositing a superconducting island with that edge shape on top of the underlying magnetic material. If the nearly-flat CMEM is indeed robust, we hypothesize that geometric deformations of the edge would preserve the scaling of number of states in the CMEM with the perimeter of the deformed edge.

We substantiate the perimeter hypothesis with an extensive analysis of a 2D square-lattice tight-binding discretization of the skyrmion model \Eq{eq: BdG_Skyrmion}, which upon setting the skyrmion exchange strength $J$ to zero outside the skyrmion edge, i.e., radius $R_{\rm sk}$, gives consistent results with the radial model (see Methods 2D tight-binding Hamiltonian, Supplementary Note 4 Counting the number of edge states of circular skyrmion and Supplementary Figure 4).

Next we consider two more geometries where the edge of the system is far from a circle and count their edge states as the overall system size is varied (see Supplementary Note 4 Defining the geometries and edge state counting and Supplementary Figure 5).

\begin{figure}
\centering
\includegraphics[scale=.8]{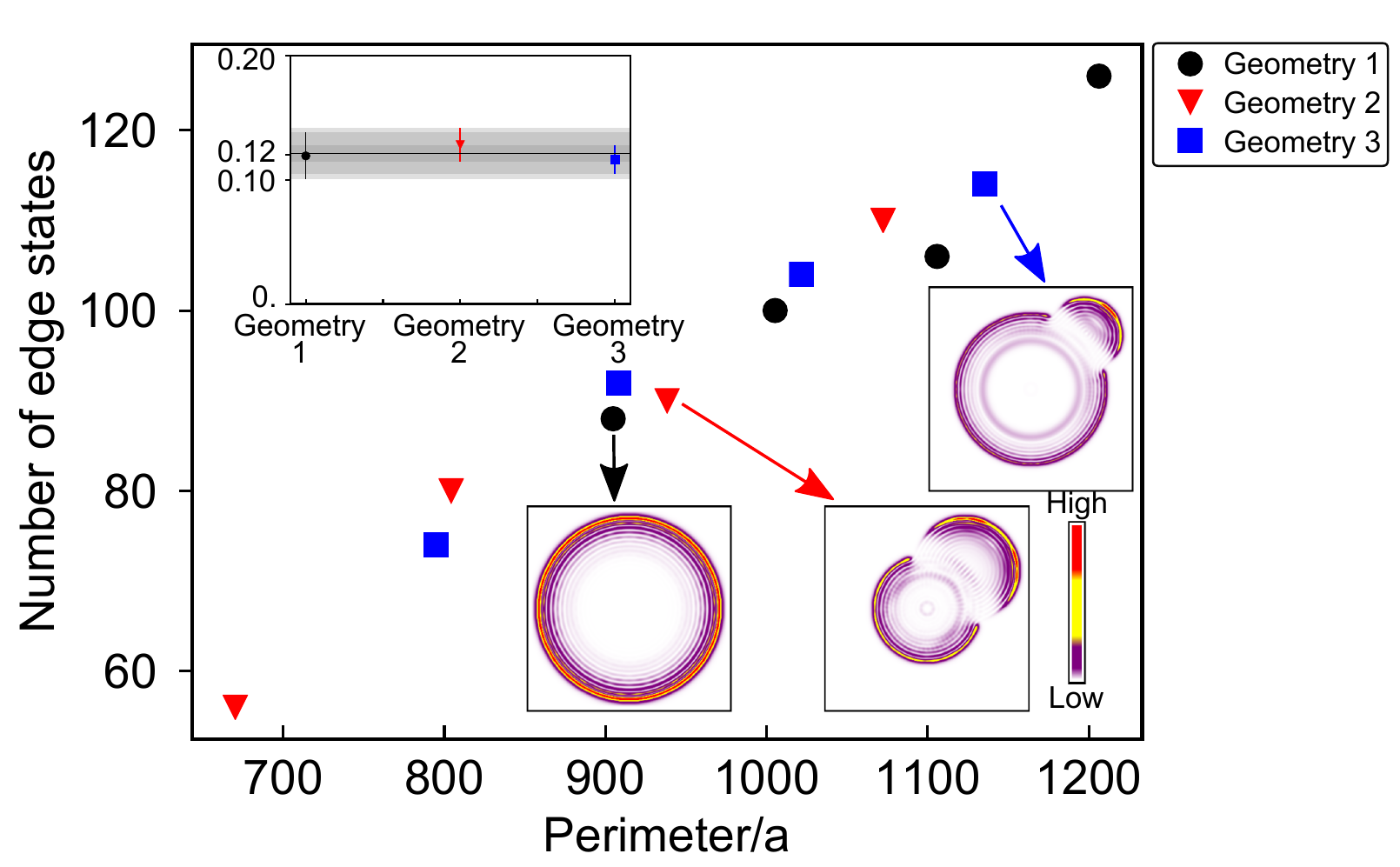}
\caption{{\bf Edge states of deformed structures.} Number of low-energy edge states as a function of the perimeter for each of the three geometries of the edge geometry 1, 2, and 3. Geometry 1 consists in a circular skyrmion while geometry 2 and 3 are made of two overlapping disks obtained from a larger skyrmion by setting the exchange field to 0 outside the desired region. The radius of the additional disk is equal to that of the central disk for geometry 2 and half this for geometry 3 (see Supplementary Note 4 Defining the geometries and edge state counting). In all geometries, the underlying skyrmion texture has azimuthal winding number $q=1$. The other model parameters are the spin-flip length $\lambda/a = 16$, the $s$-wave order parameter $\Delta_0/t = 0.1$, the chemical potential $\mu/t = 0$ and the exchange interaction strength $J/t = 0.2$. For geometry 1, the radial winding number is $p = 9$ while for geometry 2 and 3, the skyrmion has $p=6$ and $p=9$, respectively.
The graphic for black disks (geometry 1) shows the real-space image of the local density of states of one typical low-energy state. Graphics for red triangles and blue squares (geometry 2 and 3, respectively) show the local density of states averaged over the 30 lowest-energy states. The colorbar represents the local density of states from low (black) to high (red) values with different scales for the three geometries.
Top inset: linear slope extracted for each geometry. The gray shading indicates error bars from the fitting procedure. The black horizontal line is the average slope estimated to be $0.12(1) \, a^{-1}$ where $a$ is the lattice spacing.}
\label{fig: fig3}
\end{figure}

The results for the number of edge states vs. the perimeter of the edge are displayed in \Fig{fig: fig3}. It clearly shows that the number of edge states scales linearly with the perimeter of the edge for all three geometries considered, with a mean slope of $0.12(13) \, a^{-1}$. The inverse slope is a lengthscale $\xi$ associated to the edge state. We find that $\xi\approx 0.5 \lambda$ for the parameters considered, where the lengthscale $\lambda$ measures the distance for a single radial spin flip, \ie, $R_{\rm sk}= p\lambda$. This is consistent with the observed localization length of edge states in the radial direction.
This typical radial width of the edge states thus ranges from a few nanometers for the skyrmions depicted in Fig. \ref{fig: fig3} to a few tens of nanometers for the skyrmion depicted in Fig. \ref{fig: fig1}.

To further investigate the robustness of the states forming the single nearly-flat CMEM, we notice that the states in the CMEM seem to locally hybridize where the shape of the edge has sharp features. Sharp features in the edge shape allow the edge-state wavefunctions to overlap as they decay perpendicularly to the edge. Therefore the ``elastic perimeter law'' demonstrated in \Fig{fig: fig3} is best exhibited when the curvature of the edge is constant on lengthscales comparable to the extent of a single edge-state wavefunction $\xi$, as we additionally confirm through an investigation of elliptical skyrmion geometries (see Supplementary Note 4 Elliptic geometry and Supplementary Figure 6). 

Sharp corners in the shape of the edge seem a stronger perturbation than uncorrelated scalar disorder, since we numerically show using the 2D tight-binding model that the nearly-flat CMEM is indeed robust to uncorrelated scalar disorder (see Supplementary Note 5 and Supplementary Figure 7).

\section{Discussion}
In summary, we have shown that a system composed of a magnetic skyrmion coupled to a conventional $s$-wave superconductor realizes a topological superconducting phase with a nearly dispersionless chiral Majorana mode at its edge. Deforming the edge of the skyrmion away from a circular shape shows that the number of edge states can be tuned and scales linearly with the perimeter of the edge.

As skyrmions usually appear in ferromagnetic thin films, we also considered the effect of a ferromagnetic background on the edge states. For this purpose, in the radial tight-binding model we move the boundary of the system farther than the edge of the skyrmion, filling the added space with a ferromagnetic exchange field without changing the strength of the interaction $J$. We find that the edge states initially localized at the edge of the skyrmion delocalize in the background, as seen in Supplementary Note 6 and Supplementary Figure 8. This can be understood rather simply because the superconductor is gapless in that region. The delocalization of the edge states is consistent with the analytical treatment of Yang et al \cite{Loss_Majorana_skyrmion}.  

Our analysis is carried out in the Bogoliubov-de Gennes formalism without self-consistency. We believe that a self-consistent calculation would not change our main conclusions since self-consistent calculations\cite{Bjornson_selfconst_2015, Christen2016} on similar systems related to one-dimensional wire Hamiltonians, to which we map the skyrmion, didn't show any qualitative change of the physics. The only effects would then be expected near the topological phase transition where the gap is small. In our case, this may for example slightly shift the value of $\mjs$ defined in Eq.~\eqref{eq: critical_m} at which the gap closes. Furthermore, we assumed that our skyrmion arises in a magnetic insulator which guarantees that the mutual interplay between the magnetic insulator and the superconductor is weak. 

The chirality of our CMEM is determined by the azimuthal winding number $q$, so the questions arise whether the CMEM appears in other textures that have azimuthal winding, and whether chiral materials are necessary. First, note that our texture definition in \Eq{eq: skyrmion_def} may describe both Bloch and N\'{e}el skyrmions by adding a constant phase shift, named ``helicity''\cite{TokNag2013} that can be unitarily removed from our model and does not affect the discovered spectrum nor the wavefunction localization. 
Second, we have so far focused on skyrmions but our findings also apply to magnetic bubbles that have a different microscopic stabilizing mechanism but have the same topology\cite{Kiselev:2011, Tchernyshyov:2012, Buttner:2018, BernandMantel:2018}. 
For our purpose, the key distinguishing aspect of bubbles is their spatial profile: bubbles are essentially annulus-shaped domains of uniform polarization separated by ring-shaped domain walls\cite{Kiselev:2011, Tchernyshyov:2012, Buttner:2018, BernandMantel:2018}. We include this spatial feature of a bubble directly in our exchange field model Eq.~\eqref{eq: skyrmion_def} by tuning the function $f(r)$, and we show in Supplementary Note 7 and Supplementary Figure 9 and 10 that our main results, \ie, the gapped spectrum with a CMEM and wavefunction localization, appear in the bubble model too. This indicates that a wider spectrum of materials and textures could be experimentally explored for realization of our predictions.

The realization of topological superconductivity and the edge states in our system puts constraints on the parameter values. We consider three requirements for successful realization: (i) A substantial effective $p$-wave gap in the $m_{\rm J} = 0$ sector, \eg $\Delta_{\rm eff}/\Delta_0>5\%$. An estimate of the effective gap\cite{AliceaPRB} based on the skyrmion-induced spin-orbit coupling and chemical potential $\mu\left(m_{\rm J} \right) =  \mu -  \left(m_{\rm J}^2+\frac{q^2}{4}\right)/\Rsk^2 \approx \mu$ when $\Rsk$ large, is given by
\begin{align}
\Delta_{\rm eff} = \frac{\pi}{\lambda} \frac{\Delta_0}{J} \sqrt{J + \mu}
\label{eq: effective_gap}
\end{align}
where $\lambda = \Rsk/p$ is the spin-flip length. In this formula, all energy scales are in units of hopping energy $t$ which determines the bandwidth and we take it to be $t \sim 1 \, {\rm eV}$; the unit lengthscale in the formula is the lattice spacing $a$ whose dimensionful value should correspond to the microscopic electron lengthscale, so we take $a \sim 0.1\,{\rm nm}$. The requirement (i) now says that the exchange strength cannot be too large, \ie $J/t < 600 \, a/\Rsk$, assuming $p\lesssim10$. Since in materials generically $J \sim 1-10\,{\rm meV}$, the skyrmion size is allowed to reach micrometers. The second requirement is that: (ii) The topological regime is reached, so that the exchange scale $J$ surpasses the superconducting pairing $\Delta_0$. This means $\Delta_0$ is below the $1-10\,{\rm meV}$ range, or the coherence length is in the range $10-100\,{\rm nm}$, which is generally realistic. The final requirement is that: (iii) The CMEM is localized at the edge, i.e., the localization length of the edge-state wavefunctions (estimated to be $a\cdot t/\Delta_{\rm eff}$) has to be an order of magnitude smaller than the skyrmion radius $R_{\rm sk}$. From Eq.~\eqref{eq: effective_gap} using $J\approx\Delta_0$ we get the constraint that radial winding $p \sim 10$, consistent with Yang et al \cite{Loss_Majorana_skyrmion}.  One may try to relax this constraint by increasing the exchange strength.

For the superconducting part of our setup, we propose aluminum which is a known superconductor and has negligible atomic spin-orbit coupling, in accord with our general assumption that Rashba and intrinsic spin-orbit couplings are sufficiently weak. First of all, disordered thin films of aluminium have a critical temperature of the order of 3K with a coherence length of around 50 nm\citep{Meservey:1971, Dressel:2008} or less, which is within our theoretically relevant range. Second, direct measurements of the Rashba spin-orbit coupling in such thin films are hard to come by, but we find an estimate of $E_{\rm so}/\Delta_0=5\%$ for the ratio of energy scale $E_{\rm so}$ of spin-orbit scattering to the energy scale $\Delta_0$ of $s$-wave pairing in thin-film aluminum\cite{Meservey:1994}.
The skyrmion-induced spin-orbit energy scale $E_{\rm so}=(\pi^2/2m)\cdot(p^2/R^2)$ [\cf derivation of the wire model, Eq.~\eqref{eq: BdG_Hamiltonian_cont_Cylinder_wire}] can without problem reach $5\%\Delta_0$ or several times higher for theoretically relevant values of the parameters $p\lesssim 10$, $R_{\rm sk}\sim 10 - 100\, {\rm nm}$ and $\Delta_0\sim 1 \,  {\rm meV}$.

For the experimental realization of our findings we propose that the magnetic material be insulating so as to protect the CMEM. From the materials perspective, there are currently two known insulators hosting skyrmions, ${\rm Cu}_2{\rm O}{\rm Se}{\rm O}_3$ and ${\rm Ba}{\rm Fe}_{12-x-0.05}{\rm Sc}_x{\rm Mg}_{0.05}{\rm O}_{19}$ ($x = 1.6$) \cite{Seki2012, Tokura2012, White2012}. In terms of their parameters, 100 nm-thick ${\rm Cu}_2{\rm O}{\rm Se}{\rm O}_3$ films host skyrmions of radius 25 nm at temperatures ranging from a few Kelvins up to 57 K\cite{Everschor-Sitte:2018}. There is a sizeable electronic gap of 2.5 eV at 15 K\cite{Versteeg:2016}, while the lattice constant is 8.925 $\AA$ \cite{Portnichenko:2016}. All these parameters are within the ranges for which our results are relevant, as detailed in the previous paragraphs. We note that in ${\rm Ba}{\rm Fe}_{12-x-0.05}{\rm Sc}_x{\rm Mg}_{0.05}{\rm O}_{19}$ ($x = 1.6$) the skyrmions are larger, but could be within the upper limit of the tens-of-nanometers range we consider for this parameter.
If these magnetic insulators could be grown on a metallic substrate, then one may consider a finite superconducting island deposited on top of the system making the system suitable for Scanning Tunneling Microscopy/Spectroscopy (STM/STS) experiments.
Additionally, our model might also apply to the case of a metallic magnet, although feedback effects between the texture and the electrons (not considered here) can be important \cite{Motome2018, NogueraPRB2018}. In that regard, we note that skyrmions displaying a three-ring structure where observed experimentally, albeit with a change in the helicity \cite{Tokura2012}. Further, magnetic skyrmions with $q = 2$ have also been predicted in frustrated\cite{MostovoyNC2015} and itinerant\cite{MotomeLargeQ2017} magnets. An alternative platform to consider would be thick permalloy (${\rm Ni}_{81}{\rm Fe}_{19}$) disks\cite{TargetSK-permalloy2018}, since the existence of skyrmions with $p$ up to 3 was recently shown in them, although this would require a different setup. High-$p$ skyrmions were also recently observed in Pd/Fe/Ir(111) magnetic islands\cite{Wiesendanger_target_exp}. These systems, albeit metallic, naturally provide an edge to localize the CMEM and remove the need to grow a superconducting island. These results are important developments since the larger $p$ also ensures the localization of the CMEM.

The biggest challenge in the experimental verification of our findings lies in the choice of the materials. Indeed, both ingredients (skyrmions and superconductivity) are separately well-controlled and well-understood, but little is known about their combination.  In particular, we expect that the strength of the exchange field will depend on the achieved interfacing between the magnetic and superconducting materials, which is hard to predict.
Recent works aiming at engineering topological superconductivity by using magnetic adatoms or external magnetic fields have shown interesting possibilities, which means that bringing together the magnetism/spintronics and topological superconductivity communities holds great promises.

{\bf Acknowledgement:} \\
We would like to acknowledge useful conversations with Marco Aprili, Freek Massee, Stanislas Rohart, Nicholas Sedlmayr and Silas Hoffman. This work has been partially supported by French Agence
Nationale de la Recherche through the contract ANR
Mistral.\\

{\bf Author contributions:} \\
M. G. performed the numerical and analytical calculations under the supervision of A. M. and P. S. All authors discussed the results and contributed to the final manuscript.\\

{\bf Data availability:} \\
Codes and datasets used in this study are available from the corresponding author upon reasonable request.\\

{\bf Competing interests:} \\
The authors declare that they have no competing interests.\\

\newpage

\begin{center}
{\bf \Large Methods}
\end{center}
\setcounter{figure}{0}
\setcounter{section}{0}
\renewcommand\thefigure{\thesection.\arabic{figure}}   
\renewcommand\appendixname{Methods}

%\section{Rotational symmetry and discretization: 1D radial tight-binding}
\section{Radial tight-binding model}
\label{app: detail_rot}

The rotational symmetry of the problem can be exploited by defining the total angular momentum operator $J_z$ around the $z$ axis perpendicular to the plane of motion of the electrons. In polar coordinates $\left(r, \theta\right)$, it is defined as $J_z = L_z + \frac{q}{2} \sigma_z$ where $ L_z = -i\partial_\theta$ is the orbital angular momentum. Denoting the eigenvalues of $J_z$ as $m_{\rm J}$, we can expand the electronic field operators as
\begin{align}
\psi_\sigma\left({\bf r}\right) = \sum_{m_{\rm J} = -\infty}^{+ \infty} e^{i\left[m_{\rm J} - \frac{q}{2}\left(\sigma_z\right)_{\sigma \sigma}\right]\theta} \widetilde{\psi}_{m_{\rm J}, \sigma}\left(r\right)
\end{align}
%were $\widetilde{\psi}_{m_J, \sigma}\qty(r)$ annihilates an electron with spin $\sigma$ and angular momentum $m_J$ at position $r$. 
The Nambu spinor $\Psi \left({\bf r}\right)$ can thus be expanded as
\begin{align}
\Psi\left({\bf r}\right) = \sum_{m_{\rm J} = -\infty}^{+ \infty} e^{i\left[m_{\rm J} - \frac{q}{2}\sigma_z\right] \theta}\widetilde{\Psi}_{m_{\rm J}}\left(r\right)
\end{align}
where $\widetilde{\Psi}_{m_{\rm J}}\left(r\right) = \left(\widetilde{\psi}_{m_{\rm J}, \up}\left(r\right), \widetilde{\psi}_{m_{\rm J}, \down}\left(r\right), \widetilde{\psi}^\dagger_{-m_{\rm J}, \down}\left(r\right), -\widetilde{\psi}^\dagger_{-m_{\rm J}, \up}\left(r\right)\right)^T$.
We conveniently rescale the spinor by $\sqrt{r}$ so that the $r dr d\theta$ measure simplifies to $drd\theta$. After all these transformations, the BdG Hamiltonian is block-diagonal in angular momentum space and a single block $\widehat{\H}_{m_{\rm J}}\left(r\right)$ reads
\begin{widetext}
\begin{align}
\widehat{\H}_{m_{\rm J}}\left(r\right) = -\frac{1}{2m}\left[\partial_r^2 +\frac{1}{4 r^2}\left(1 - q^2 - 4 m_{\rm J}^2 + 4 q m_{\rm J}\sigma_z\right) - \mu\right] \tau_z + J	\, \sigma_z \cos f + J \, \sigma_x \sin f + \Delta_0 \, \tau_x
\label{eq: BdG_rescaled_continuum}
\end{align}
\end{widetext}
We discretize the remaining polar $r$ variable by introducing a lattice spacing $a$ so that $r \to r_j = j a$, and in numerical calculations we set $a = 1$. The nearest-neighbor tight-binding Hamiltonian uses the Nambu basis ${\cal C}^\dagger_{j} =  \left(c^\dagger_\up \left(j a\right), c^\dagger_\down \left(j a\right),c_\down \left(j a\right), -c_\up \left(j a\right) \right)$. %where we have used the notations $c$ and ${\cal C}$ for fields and spinors respectively instead of $\psi$ and $\Psi$ to emphasize their discrete nature.
We parametrize the tight-binding Hamiltonian as
\begin{align}
\begin{split}
\widehat{H}_{m_{\rm J}}^{\rm TB} = \sum_{j = 1}^L {\cal C}^\dagger_{j+1} \, M \, {\cal C}_j & + \hc   + {\cal C}^\dagger_j \, C \, {\cal C}_j
\end{split}
\label{eq: TB_form}
\end{align}
Now, we Taylor expand \Eq{eq: TB_form} to second order, integrate by parts and identify the matrices $M$ and $C$ from \Eq{eq: BdG_rescaled_continuum}.
This leads to
\begin{align}
\begin{split}
\widehat{H}_{m_{\rm J}}^{\rm TB} \approx \sum_{j = 1}^L -t \, {\cal C}^\dagger_{j+1} \tau_z {\cal C}_j & + \hc   + {\cal C}^\dagger_j \left[2t - \mu - \frac{t}{4 j^2}\left(1-q^2 - 4 m_{\rm J}^2 + 4 q m_{\rm J} \sigma_z\right)\right]\tau_z {\cal C}_j  \\ & + {\cal C}^\dagger_j \left[J \sigma_z \cos f + J\sigma_x \sin f + \Delta_0 \tau_x\right] {\cal C}_j.
\end{split}
\label{eq: Block_SP_Hamiltonian_TB}
\end{align}
The tight-binding hopping energy $t = 1/\left(2m a^2\right)$ in terms of the effective electron mass $m$.
We exactly diagonalize the Hamiltonian in the form of \Eq{eq: Block_SP_Hamiltonian_TB} without implementing the self-consistency inherent to the Bogoliubov-de-Gennes formalism.

\section{2D tight-binding Hamiltonian}
\label{app: detail_2D}

On the square lattice ${\bf r} = \left(x a, y a\right)$ where $a\equiv1$ is the lattice spacing, and $x,y$ are integers labelling the sites of the lattice, the two-dimensional tight-binding Hamiltonian is
\begin{align}
\begin{split}
H^{\rm 2D \, TB} & =  \sum_{{\bf r}=x,y}\left[ \sum_{\sigma = \up, \down} -t \,c^{\dagger}_{{\bf r} +{\bf \hat{x}}\sigma}c_{{\bf r}\sigma} -t\, c^{\dagger}_{{\bf r} + {\bf \hat{y}}\sigma}c_{{\bf r}\sigma}  + (4t - \mu) \, c^{\dagger}_{{\bf r}\sigma}c_{{\bf r}\sigma} \right.\\ &+ \Delta_0 \, c^{\dagger}_{{\bf r} \up}c^\dagger_{{\bf r} \down} + \hc
\\ &\left. + J\,\sum_{\sigma,\sigma'}  c^{\dagger}_{{\bf r} \sigma}\left({\bf n}\left({\bf r}\right) \cdot {\bm \sigma}\right)_{\sigma \sigma '}c_{{\bf r} \sigma'}\right],
\end{split}
\end{align}
where the parameters are the same as in the main text, and $t$ is the hopping amplitude, $\mu$ the chemical potential measured from the bottom of the band, $\Delta_0$ the $s$-wave gap and $J$ the exchange coupling with the texture. The unit vector in the $x$ (resp. $y$) direction is denoted as ${\bf \hat{x}}$ (resp. ${\bf \hat{y}}$). Exact diagonalization is then performed without implementing the self-consistency inherent to the Bogoliubov-de-Gennes formalism.

Consistently with the radial model, in the regime of $J/\Delta_0$ large enough (estimated as $J>\sqrt{\Delta_0^2+\mu^2}$) and for any $q$ we find weakly-dispersing states extended around the edge and localized in the radial direction near $R_{\rm sk}$, while only for $q$ even there is a zero energy state localized at the skyrmion center.

\section{Gradient and Laplace operators in the cone geometry}
\label{app: laplace}
%or just Cone geometry?
As in Wu et al\cite{MappingMartin}, consider a cone of half-opening angle $\varphi$ and base radius $\Rsk$ where the coordinates $r$ and $\theta$ respectively denote the distance measured from the tip of the cone and the usual polar angle. Denoting by ${\bf \hat r}$ and $\hat {\bm \theta}$ the unit vectors on the cone, the gradient and Laplace operators read
\begin{align}
& {\bm \nabla} = \partial_r \,{\bf \hat r}+ \dfrac{1}{r \sin \varphi} \partial_\theta \, \hat {\bm \theta} & \nabla^2 =  \partial_r^2 + \dfrac{1}{r}\partial_r + \dfrac{1}{r^2 \sin^2 \varphi} \partial_\theta^2.
\end{align}
The cylinder limit is $\varphi = 0$, $r \to \infty$ while keeping $r \sin \varphi = {\rm const} = \Rsk$. Under the transformation from the disk to the cylinder via the cone, the surface element varies like
\begin{align}
r \, dr \, d\theta \rightarrow r \, \sin \varphi \, dr \, d\theta \rightarrow \Rsk \, dr \, d\theta.
\end{align}
\newpage
\cleardoublepage

%\setcounter{section}{0}
%\setcounter{equation}{0}
%\renewcommand\appendixname{Supplementary Materials}
%\renewcommand\thefigure{\thesection.\arabic{figure}}   
%\renewcommand\theequation{SM \thesection.\arabic{equation}}   
%\begin{center}
%{\bf \Large Supplementary Materials}
%\end{center}
%
%\counterwithin{figure}{section}

\newpage
\bibliography{SK_arxiv_v2}

\cleardoublepage

\appendix

%\title{Supplementary Information: \\Topological superconductivity with deformable magnetic skyrmions}
%
%
%\author{Maxime Garnier}
%\email[Corresponding author: \,]{maxime.garnier1@u-psud.fr}
%\affiliation{Laboratoire de Physique des Solides, UMR 8502, CNRS,
%Universit\'{e} Paris-Sud, Universit\'{e} Paris-Saclay, 91405 Orsay, France}
%\author{Andrej Mesaros}
%\affiliation{Laboratoire de Physique des Solides, UMR 8502, CNRS,
%Universit\'{e} Paris-Sud, Universit\'{e} Paris-Saclay, 91405 Orsay, France}
%\author{Pascal Simon}
%\affiliation{Laboratoire de Physique des Solides, UMR 8502, CNRS,
%Universit\'{e} Paris-Sud, Universit\'{e} Paris-Saclay, 91405 Orsay, France}
%
%\date{\today}
%
%\maketitle

\setcounter{figure}{0}
\setcounter{equation}{0}
\renewcommand\thefigure{\arabic{figure}}   
\renewcommand\theequation{SM \thesection.\arabic{equation}}   
\renewcommand{\thesection}{\arabic{section}}

\begin{center}
{\large \bf Supplementary Information: \\Topological superconductivity with deformable magnetic skyrmions}
\\Maxime Garnier, Andrej Mesaros and Pascal Simon\\{\small \it Laboratoire de Physique des Solides, UMR 8502, CNRS, Universit\'{e} Paris-Sud, Universit\'{e} Paris-Saclay, 91405 Orsay, France}
\end{center}

\begin{flushleft}
{\large \bf Supplementary Figures}
\end{flushleft}

\begin{flushleft}
{\ \bf Supplementary Figure 1}
\end{flushleft}

\begin{figure}[h!]
\centering
\hspace*{-2.5cm}
\includegraphics[scale=1.]{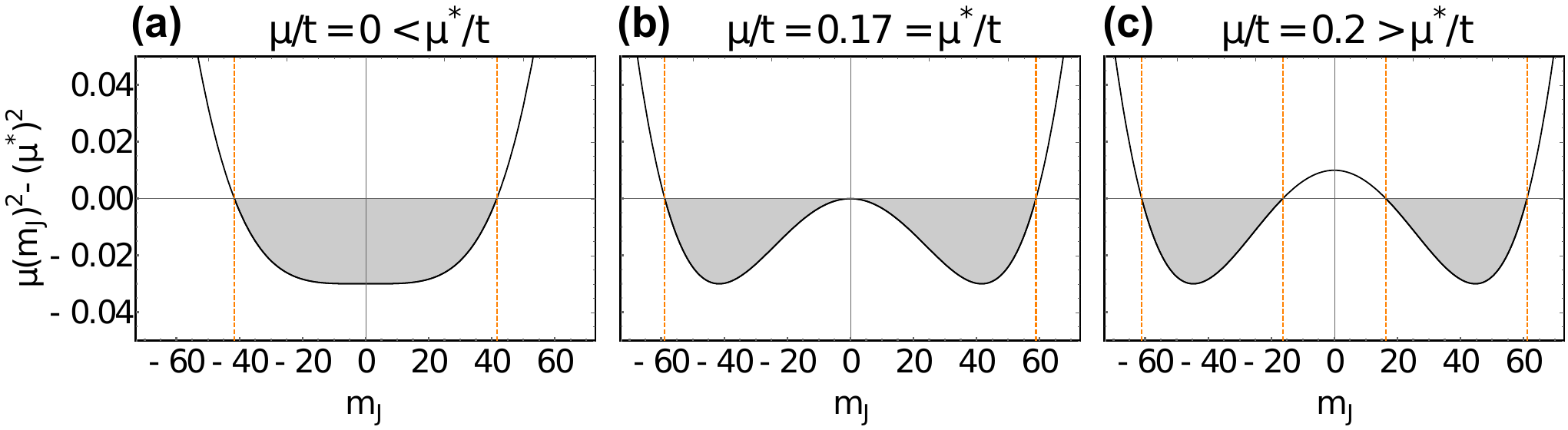}
\caption{{\bf Influence of the chemical potential on the topological properties of the wire model.} The function $\mu\left(m_{\rm J}\right)^2 - \left(\mu^*\right)^2$ for different values of the chemical potential $\mu$, in a skyrmion of radius $R_{\rm sk}/a = 100$ and azimuthal winding number $q = 2$ with parameters $\Delta_0/t = 0.1$, $J/t = 0.2$. For these parameters $\mu^{*}/t = \sqrt{J^2-\Delta_0^2} = 0.17(32)$. {\bf a} $\mu/t = 0 < \mu^*/t$, {\bf b} $\mu/t = \mu^{*}/t$ and {\bf c} $\mu/t = 0.2 > \mu^{*}/t$. The gray areas show the topologically non-trivial momentum ranges while the vertical dashed orange lines mark the gap-closing momenta.}
\label{fig: app_topo_transition}
\end{figure}

\newpage

\begin{flushleft}
{\ \bf Supplementary Figure 2}
\end{flushleft}

\begin{figure}[h!]%
\hspace*{-1.2cm}
\centering
\label{fig: slope_J0p2}
\includegraphics[scale=.4]{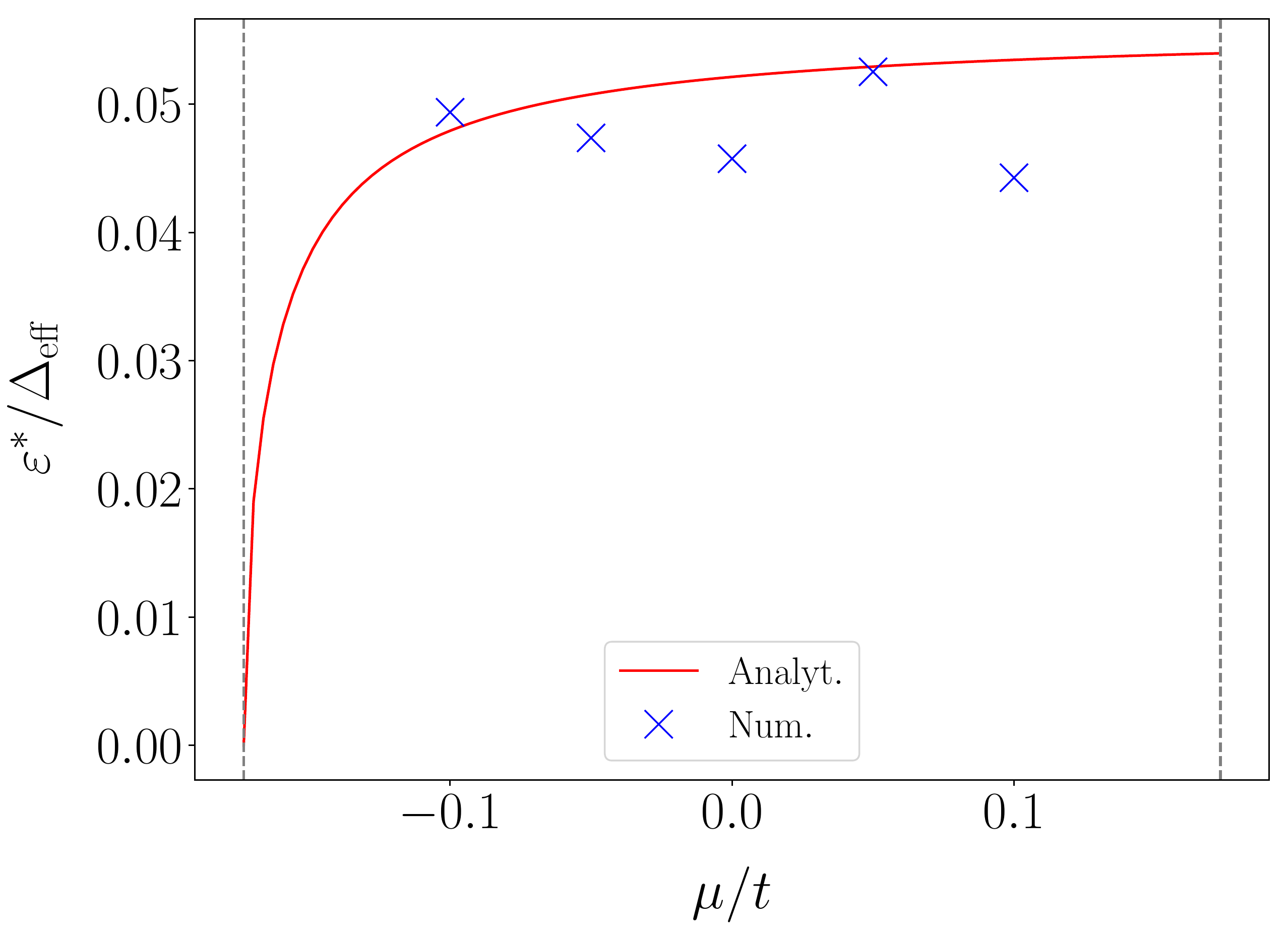}
\caption[num_check_form]{{\bf Near-flatness of the CMEM}. Ratio between the maximal energy attained by the CMEM states and the effective gap is plotted versus the chemical potential. The analytical estimate (red line) is given by Supplementary Equation~\ref{eq: slope_param} where we have chosen the prefactor of order unity to be $c \equiv 0.44$ (see text), while the numerical data (blue crosses) is obtained from the radial tight-binding skyrmion model. The parameters are $\Rsk/a = 1001$, $p=10$, $q=2$, $\Delta_0/t = 0.1$, $J/t = 0.2$. The vertical dashed gray lines mark the $\left\vert\mu\right\vert=\mus$ points.}
\label{fig: num_check_eff}%
\end{figure}

\newpage

\begin{flushleft}
{\ \bf Supplementary Figure 3}
\end{flushleft}

\begin{figure}[h!]
\centering
\hspace*{-1.2cm}
\includegraphics[scale=.6]{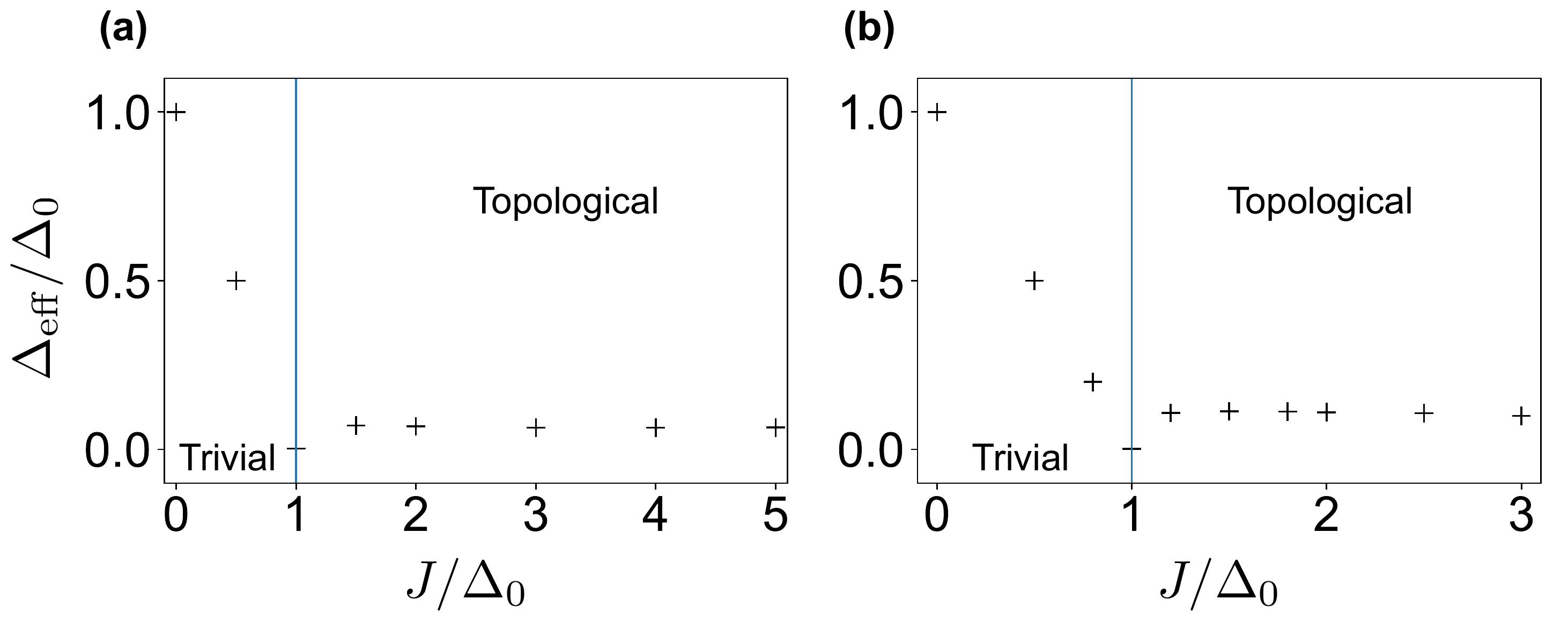}
\caption{{\bf Comparing phase diagrams of the skyrmion model on disk and the model on cylinder.} Effective gap for $\mu/t = 0$ as a function of the exchange coupling $J$ with $\Delta_0/t = 0.1$ in the radial tight-binding setup. {\bf a} Skyrmion model on the disk geometry with $p = 10$ and $R_{\rm sk}/a = 1000$. {\bf b} The mapped model on the cylinder with aspect ratio 1 and $R_{\rm sk}/a = 500$. In both {\bf a} and {\bf b}, the vertical blue line indicates the theoretical gap-closing point $J = \sqrt{\Delta_0^2 + \mu^2}$.}
\label{fig:PhDiag}
\end{figure}

\newpage

\begin{flushleft}
{\ \bf Supplementary Figure 4}
\end{flushleft}

\begin{figure}[h!]
\hspace*{-.7cm}
\centering
\includegraphics[scale=.7]{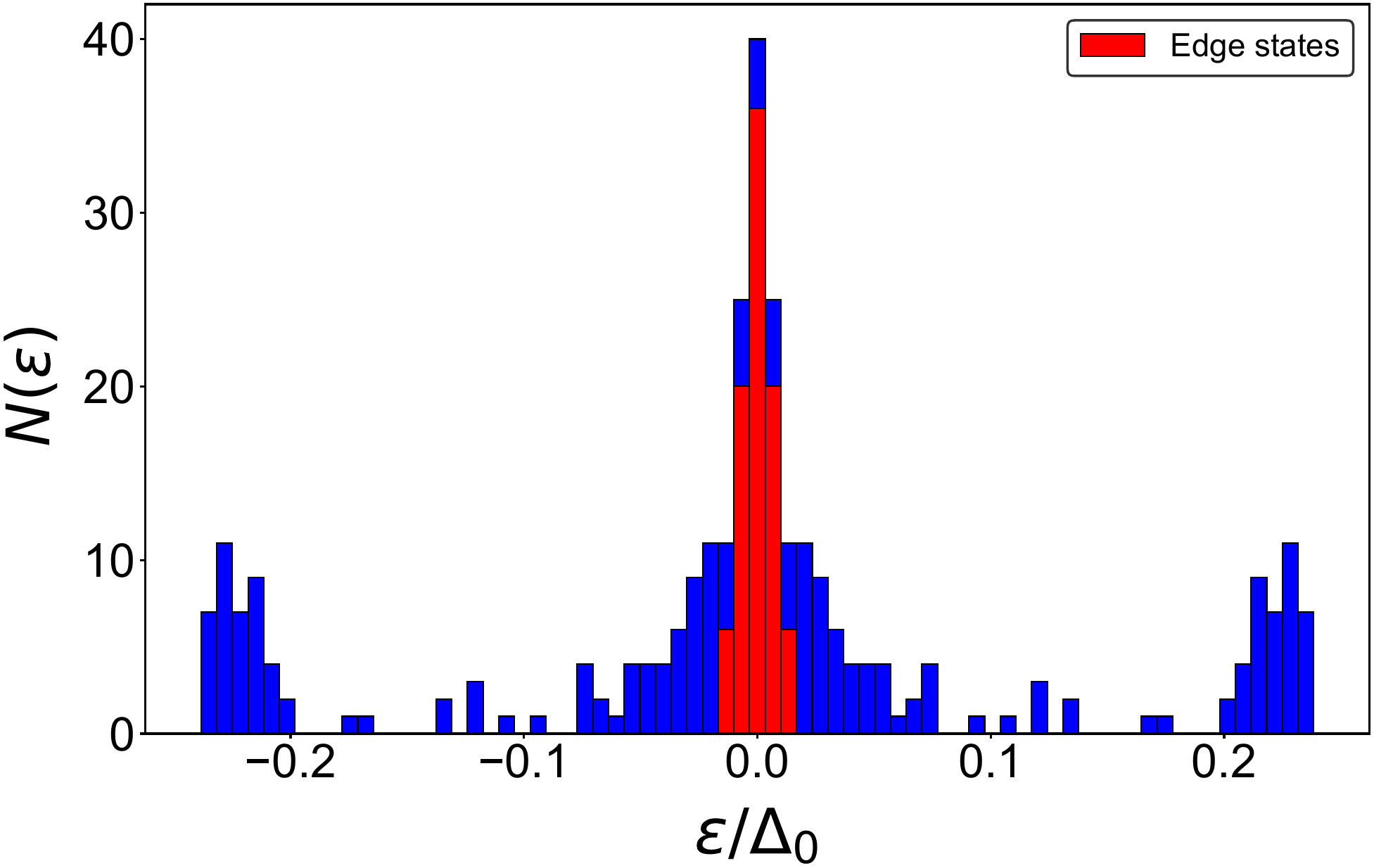}
\caption{{\bf Counting the number of edge states.} Density of states at low energies for $p = 9$ and $q = 1$ circular skyrmion with $\lambda/a = 16$. The other parameters are $\mu/t = 0$, $\Delta_0/t=0.1$, $J/t = 0.2$. The blue histogram represents the 300 lowest energy states, while the red histogram counts only the edge states among them (see definition in text, the selection criterion uses width $l = 3a$ which yields $l/\lambda \approx 0.19$). In this case there are 88 edge states. The coherence peaks in the density of states are clearly seen around $\left\vert\varepsilon/\Delta_0\right\vert \approx 0.2$. The result is consistent with an edge mode that very weakly disperses around zero energy. Other in-gap states, located in the bulk of the skyrmion, are attributed to impurity states (see main text).}
\label{fig: counting_dos}
\end{figure}

\newpage

\begin{flushleft}
{\ \bf Supplementary Figure 5}
\end{flushleft}

\begin{figure}[h!]
\centering
\includegraphics[scale=.3]{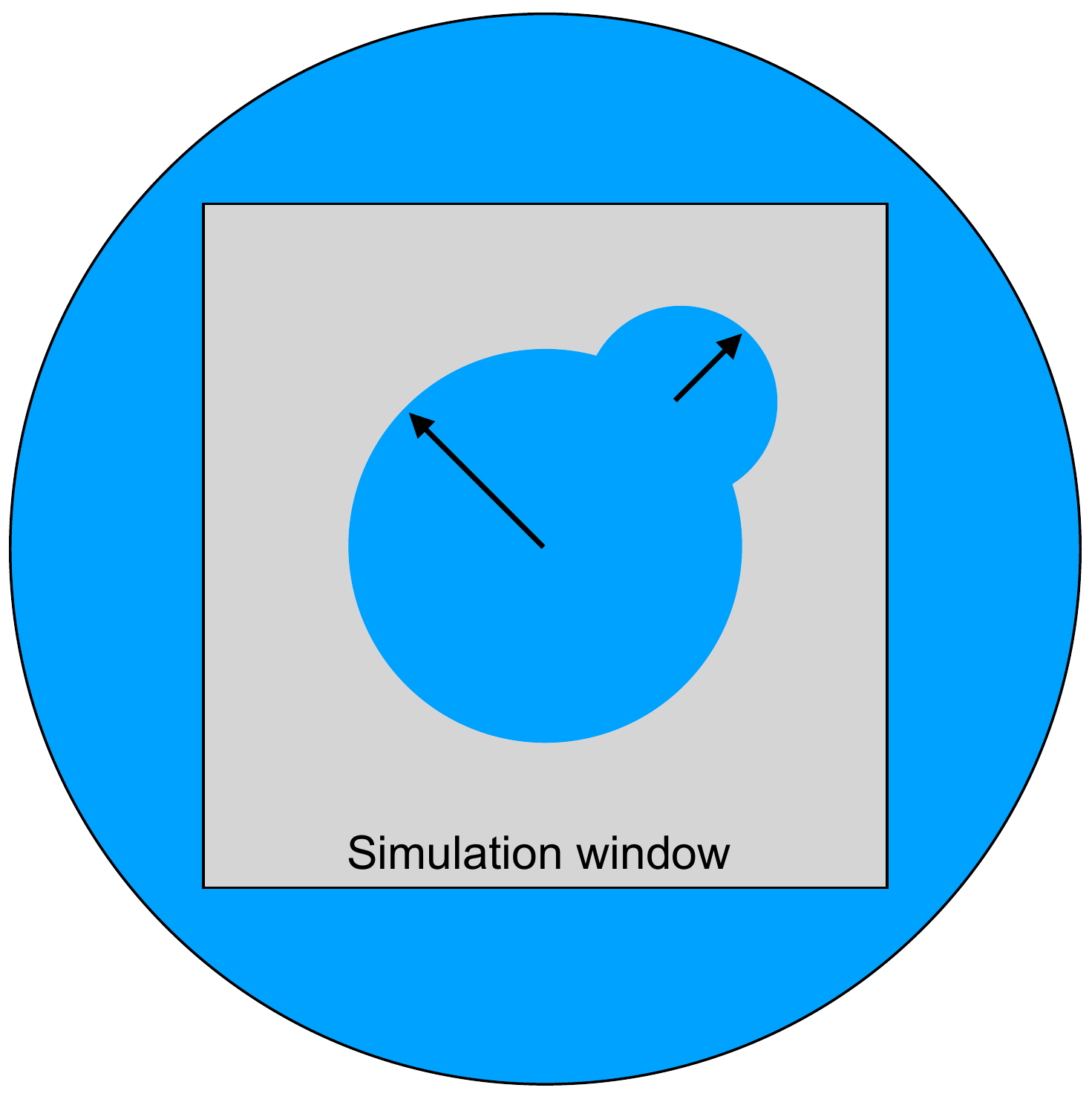}
\caption{{\bf Defining geometries 2 and 3.} The largest blue disk (extending to outmost circle) represents a temporary large skyrmion texture. We define a texture with geometry 2 or 3 by setting the exchange $J=0$ outside the inner blue region. The gray square shows the entire system kept in the calculation. The black arrows indicate the tunable radii, $r = R$ for geometry 2 and $r = R/2$ for geometry 3, where $R$ is the radius of the central circle.}
\label{fig: app_blobdef}
\end{figure}

\newpage

\begin{flushleft}
{\ \bf Supplementary Figure 6}
\end{flushleft}

\begin{figure}[h!]
\centering
\hspace*{-1.5cm}
\includegraphics[scale=1.2]{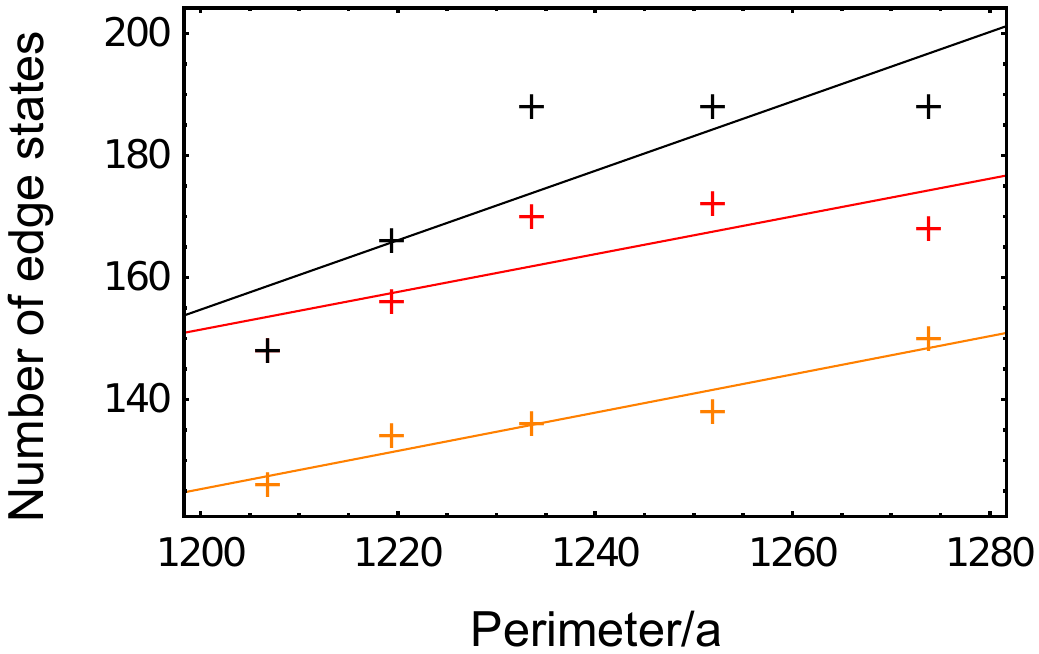}
\caption{{\bf Counting the edge states in elliptical geometry}. Result of the counting of the edge states as a function of the perimeter in elliptical geometries for 3 different edge state selection criteria $l/\lambda = 0.25$ (orange), 0.33 (red) and 0.42 (black). Solid lines represent the best linear fit. The perimeter is varied by changing the aspect ratio of the ellipse while keeping its surface area fixed. The estimated slopes are $0.313\, a^{-1}$, $0.310\, a^{-1}$ and $0.569\, a^{-1}$, respectively.}
\label{fig: ellipse_counting}
\end{figure}

\newpage

\begin{flushleft}
{\ \bf Supplementary Figure 7}
\end{flushleft}

\begin{figure}[h!]
\hspace*{-2.2cm}
\centering
\includegraphics[scale=1.]{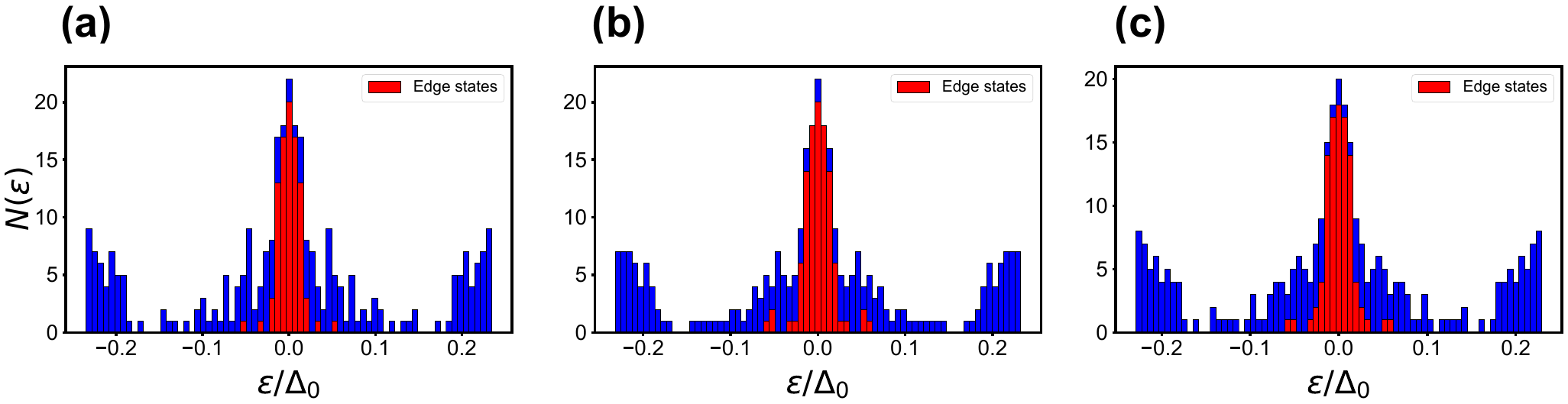}
\caption{{\bf Robustness of the CMEM to scalar disorder in the 2D tight-binding model}. Starting from the same parameters as in Supplementary Figure 5 except that $q = 2$. We implement uncorrelated scalar disorder 
%on the edge 
by randomly varying the on-site energy (see text). The variations are drawn from a normal distribution of mean $\mu$ and standard deviation $\sigma_\mu/\Delta_0 = 0.2, \, 0.5, \, 0.7$ for {\bf a}, {\bf b} and {\bf c}, respectively.}
\label{fig: app_disorder}
\end{figure}

\newpage 

\begin{flushleft}
{\ \bf Supplementary Figure 8}
\end{flushleft}

\begin{figure}[h!]
\hspace*{-.9cm}
\centering
\includegraphics[scale=.5]{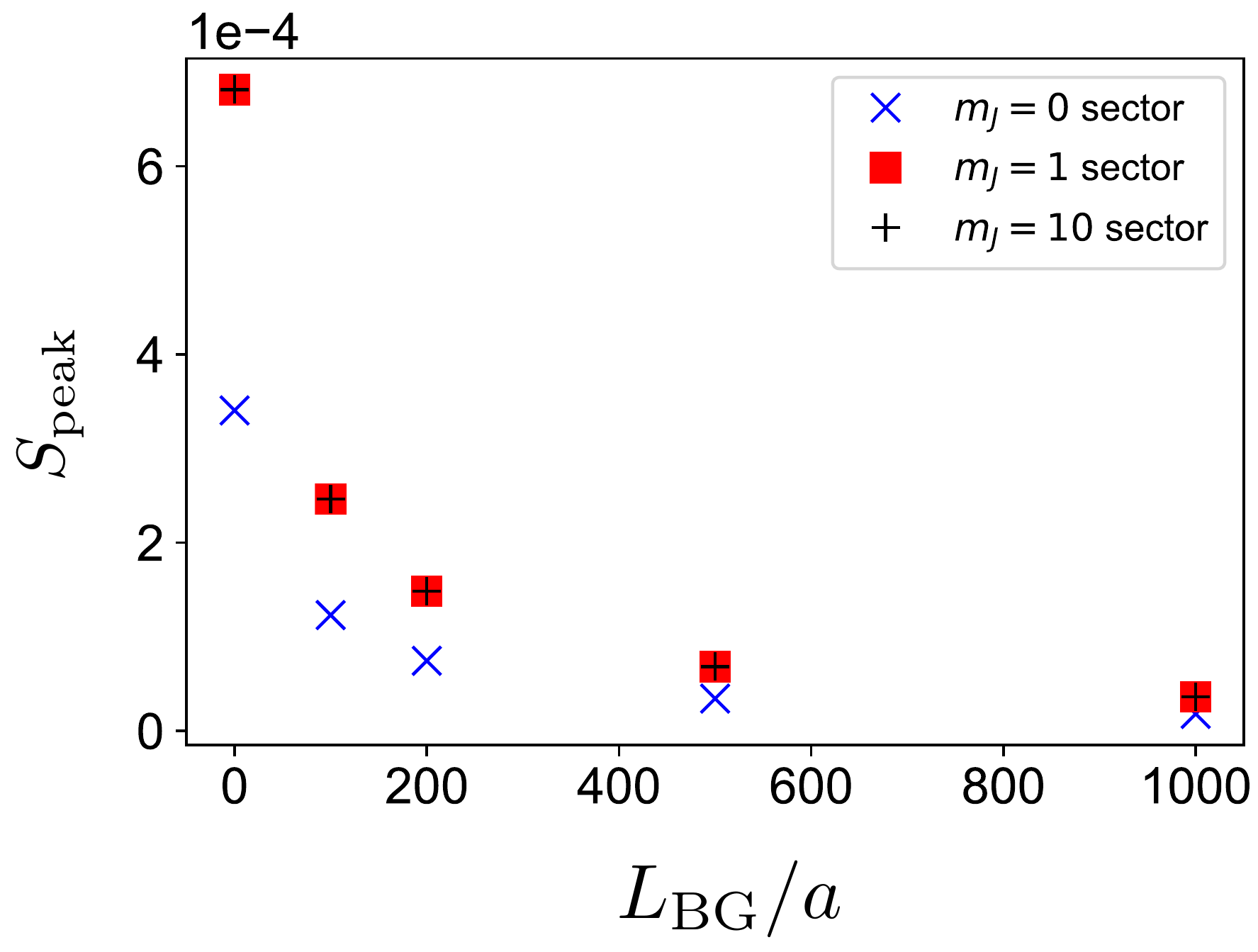}
\caption{{\bf Delocalization of the edge states in a ferromagnetic background.} Scaling of the spectral weight $S_{\rm peak}$ under the peak of edge states as function of the background size $L_{BG}$ for $L_{BG}/a \in \left\{0, 100,500,1000\right\}$ for $W_{\rm peak}/a \approx 200$ (from 800 to 1000) in the angular momentum sectors $m_{\rm J} = 0, \, m_{\rm J} = 1$ and $m_{\rm J} = 10$. The delocalization phenomenon holds in all sectors with the same decay length. The parameters used for the computation are $p = 10$, $R_{\rm sk}/a = 1000$, $\Delta_0/t = 0.1$, $\mu/t = 0$ and $J/t = 0.2$.}
\label{fig: FSS_deloc}
\end{figure}

\newpage

\begin{flushleft}
{\ \bf Supplementary Figure 9}
\end{flushleft}

\begin{figure}[h!]
  \centering
  \hspace*{-1.3cm}
\includegraphics[scale=.9]{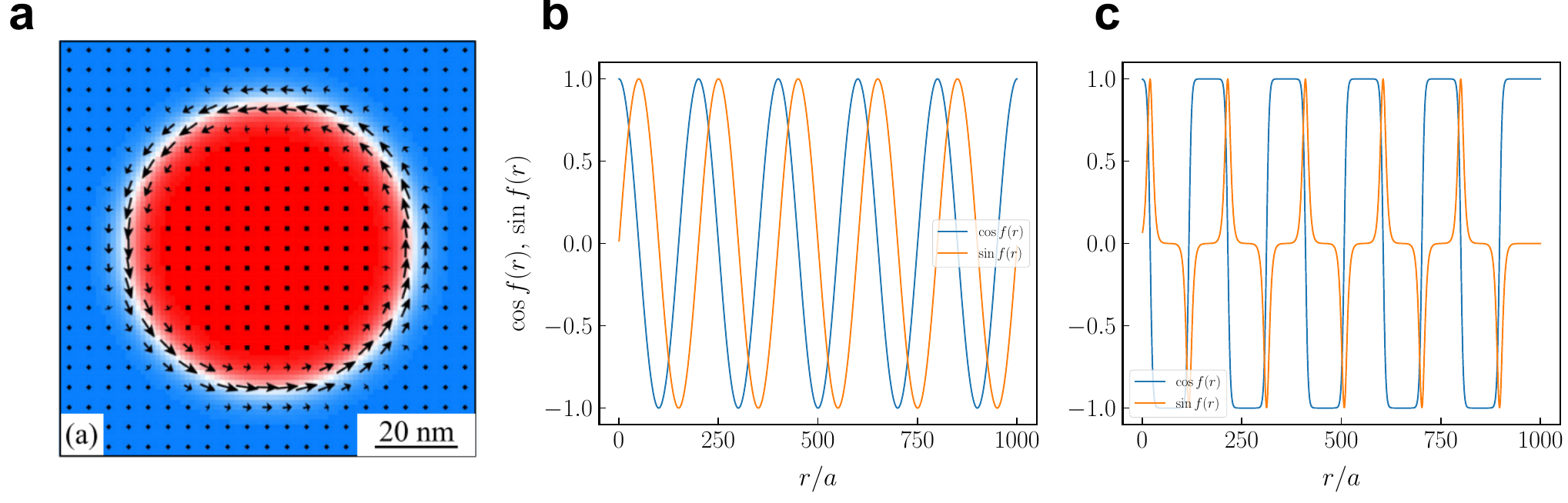}
\caption{{\bf a} Real-space profile of a bubble. The arrows represent the in-plane components of the magnetization $\left(n_x, n_y\right)$ while the color indicates the $z$ component of the magnetization from $-1$ (red) to $+1$ (blue). Reprinted figure with permission from I. Makhfudz, B. Kr\"{u}ger and O. Tchernyshyov, \emph{Physical Review Letters} {\bf 109}, 217201 (2012). Copyright (2012) by the American Physical Society. Our model of $p = 10$ {\bf b} skyrmion and {\bf c} bubble profiles. For the bubble the width of a domain wall is $w/a = 5$ (see Supplementary Equation~\ref{eq: bubbleprof}).}	
\label{fig: Sk_v_bubble_numprofiles}
\end{figure}

\newpage

\begin{flushleft}
{\ \bf Supplementary Figure 10}
\end{flushleft}

\begin{figure}[h!]
\centering
\hspace*{-1.2cm}
\includegraphics[scale=1]{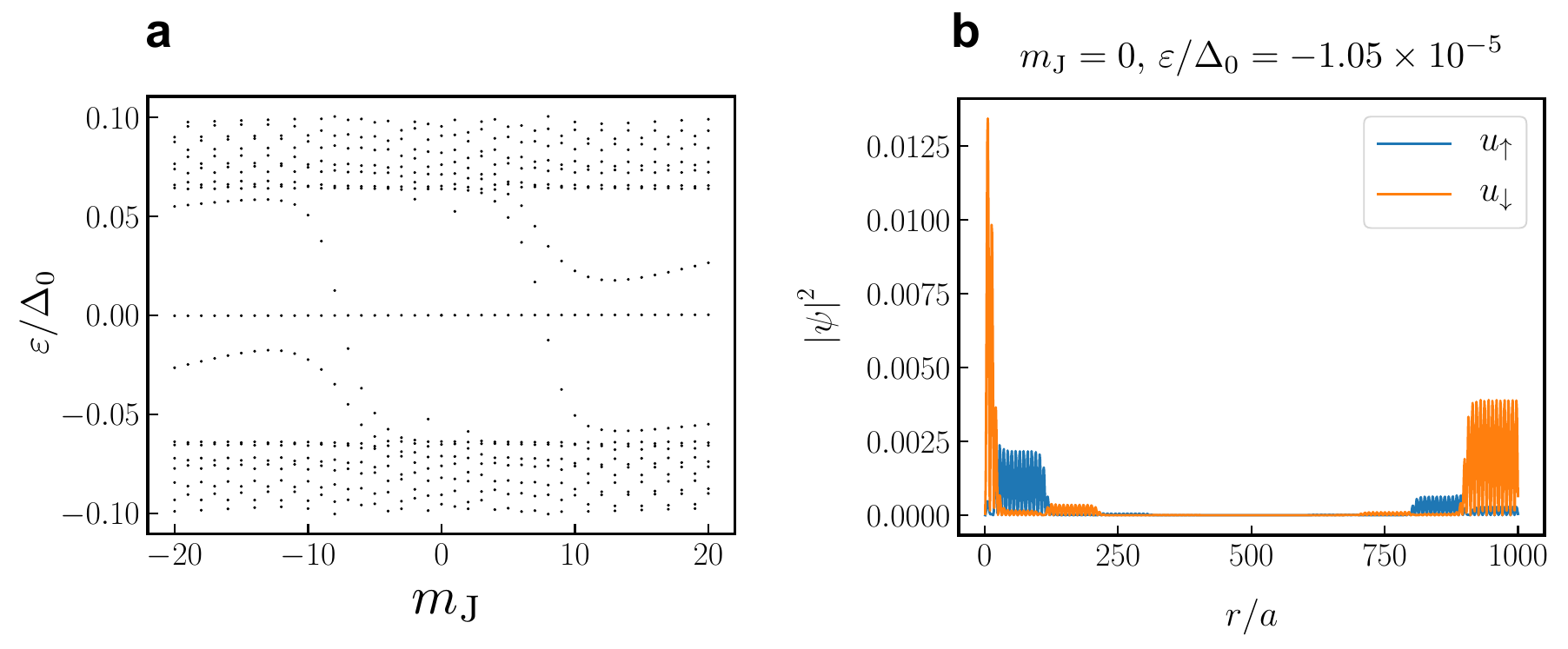}
\caption{Numerical diagonalization results for the parameters $p = 10$, $L = 1000$, $q = 2$, $J/t = 0.2$, $\Delta_0/t = 0.1$, $\mu/t = 0$ and a DW width $w/a = 5$. {\bf a} Spectrum. {\bf b} Density of one of the Majorana zero-modes.}	
\label{fig: res_w_5_p_10}
\end{figure}

\cleardoublepage

\begin{flushleft}
{\large \bf Supplementary Notes}
\end{flushleft}
\renewcommand\appendixname{Supplementary Note}

\section{Topological transitions in the wire model}
\label{app: topo_trans_reent}

The wire Hamiltonian $\H_{m_{\rm J}}^{\rm wire}(r)$ (see Eq.~(3) in the main text) corresponding to angular momentum sector $m_{\rm J}$ is in a topological phase when $\mu\left(m_{\rm J}\right)^2 - \left(\mus\right)^2 < 0$, where we introduced the quantity $\mus = \sqrt{J^2-\Delta_0^2} > 0$. This leads to two cases. Firstly, Eq.~(6)  of the main text is only valid in the case $\left\vert\mu\right\vert < \mus$, and gives the two solutions presented there. Secondly, if $\mu > \mu^*$, solving for the gap-closing momenta yields four solutions $\pm \left\vert m_{{\rm J}, \pm}^*\right\vert$, of the form
\begin{align}
\left\vert m_{J, \pm}^*\right\vert \approx R_{\rm sk}\sqrt{\left(\mu \pm \sqrt{J^2-\Delta_0^2}\right)} + {\cal O}\left(1\right).
\label{eq: critical_m_reentrant}
\end{align}
Supplementary Figure~\ref{fig: app_topo_transition} shows the different profiles of $\mu\left(m_{\rm J}\right)^2 - \left(\mu^*\right)^2$ depending on the ratio of $\mu$ and $\mu^*$, where $\mu\left(m_{\rm J}\right)\equiv\mu - \left(m_{\rm J}^2+\frac{q^2}{4}\right)/\left( 2m R_{\rm sk}^2\right)$. Supplementary Figure~\ref{fig: app_topo_transition}c shows that in the second case discussed above (not presented in the main text), the momentum range $m_{\rm J} \in \left[-m_{{\rm J},-}^*, m_{{\rm J},-}^*\right]$ is topologically trivial while the ranges $m_{\rm J} \in \left[m_{{\rm J},-}^*, m_{{\rm J},+}^*\right]$ and $m_{\rm J} \in \left[-m_{\rm{J},+}^*, -m_{\rm{J},-}^*\right]$ are non-trivial as denoted by the gray filling of the curve. Consequently, in this regime of $\mu > \mu^*$ there is no Majorana zero-mode in the core of the skyrmion even if the azimuthal winding number $q$ is even. The case $\mu < -\mus$ corresponds to the case of a fully topologically trivial system. These considerations were confirmed numerically in the radial tight-binding setup of the skyrmion model.

\section{Velocity of the chiral Majorana edge mode}
\label{app: slope}
Treating $\H_{m_{\rm J}}^{\rm slope}(r)$ in Eq.~(4) in the main text as a first order perturbation to the Majorana flat band (MFB) of the $q=0$ model, we can estimate the upper limit on the energy $\varepsilon^*$ that the CMEM reaches, which occurs at the maximal $m_{\rm J}$ of the CMEM, \ie, $\varepsilon^*\equiv \vert \varepsilon^{\rm edgestate}(\left\vert \mjs*\right\vert)\vert$. This reads
\begin{align}
\varepsilon^* = c\frac{q\, \mjs}{2 m \Rsk^2}
\end{align}
where $\mjs$ is given by Eq.~(6) in the main text, assuming the case $\left\vert \mu \right\vert < \mus$. We have also assumed $\left\langle \tau_z\sigma_z r^{-2}\right\rangle_{\rm MFB state} = c\Rsk^{-2}$, taking into account that the edge states are very localized around $\Rsk$ for the relevant range of model parameters, and expecting that $c$ is a constant of order unity.

The effective $p$-wave gap $\Delta_{\rm eff}$ being estimated by Eq.~(9) in the main text, we obtain for the ratio between the maximal CMEM energy and the effective gap:
\begin{align}
\frac{\varepsilon^*}{\Delta_{\rm eff}} = c\frac{q}{p}\frac{J}{\pi\Delta_0} \sqrt{\dfrac{\mu+\sqrt{J^ 2 - \Delta_0^2}}{J + \mu}},
\label{eq: slope_param}
\end{align}
where all energies are in units of the bandwidth $t$. We next compare this estimate to the numerical results in the radial tight-binding skyrmion model, see Supplementary Figure~\ref{fig: num_check_eff}. First, we have calculated numerically the expectation value $\left\langle \sigma_z\tau_z\right\rangle_{\rm MFB state}$ and indeed found typical values $\approx\!0.44$ of order unity. Supplementary Figure~\ref{fig: num_check_eff} shows a good agreement between our numerics and our analytical estimate for $c\equiv0.44$. Due to the large size of the skyrmion used here to minimize finite-size effects, the ratio Supplementary Equation~\ref{eq: slope_param} is of order $5 \%$. From the computation on smaller skyrmions and by varying parameters $J,\,\mu$, we find that this value can be increased to at most $\approx\!10\%$ and the CMEM therefore universally appears almost flat. In fact, our analytical estimate confirms that the near-flatness of the CMEM cannot significantly change by varying the parameters $J,\,\Delta_0$ in their respective ranges under consideration (see Discussion).

\newpage
\cleardoublepage

\section{Phase diagrams of the skyrmion model and the cylinder model}
\label{app: PhDiag}

To support the use of our mapping from the disk to the cylinder, we here plot the two phase diagrams at a fixed value of the chemical potential $\mu/t = 0$. To do so, we tune the exchange coupling $J$ and measure the effective gap, \ie the gap in the $m_{\rm J} = 0$ sector. Supplementary Figure~\ref{fig:PhDiag} clearly shows that the effective gap behaves the same way in both models, including the topological phase transition at the analytically predicted value $J = \sqrt{\Delta_0^2 + \mu^2}$. Moreover, the agreement in the topological regime is quantitative, while the small discrepancy can be accounted for by the fact that in the cylinder model we neglected the small chemical potential renormalization and the boundary term, Eq.~(5) in the main text.

\newpage
\section{Different edge geometries: counting the states in the chiral Majorana edge mode}
\label{app: geom_counting}

In this section we define the different geometries mentioned in the main text, as well as our technique for counting the edge states.

\subsection*{Counting the number of edge states of a circular skyrmion}
\label{app: counting_edge}

We firstly investigate the original skyrmion texture that, dubbed ``geometry 1''.
The circular edge of the original skyrmion texture (``geometry 1'') is close to perfectly rotationally symmetric on the 2D lattice (except for the breaking of the spatial rotation symmetry down to the square lattice's discrete one). An eigenstate of the 2D tight-binding model is defined as an edge state if the maximum of its wavefunction lies in a certain corona of width $2l$ around the edge of the skyrmion at $r=R_{\rm sk}$. The position $R_{\rm max}$ of the maximum of the wavefunction is found using the angularly-averaged wavefunction. Precisely, a state is an edge state if $\left\vert R_{\rm sk}-R_{\rm max}\right\vert \leq l$. We will comment on the chosen values for $l$ below. Based on this definition, the count of edge states for the circular skyrmion is presented in Supplementary Figure~\ref{fig: counting_dos}, where the cut-off $l$ is 3 lattice sites so that $l/\lambda \approx 0.19 $, given the skyrmion's value $\lambda/a = 16$.

\subsection*{Defining the geometries and edge state counting} 
\label{app: blob_def}

Geometries 2 and 3 are obtained by defining a skyrmion in a large simulation window, and then defining the desired texture's edge by setting the exchange $J$ to zero outside the edge curve, as shown in Supplementary Figure~\ref{fig: app_blobdef}. The shape of the edge curve is defined as the outside perimeter of two overlapping disks. The second disk, of radius $r$ is positioned so that its center lies on the perimeter of the first disk, which has radius $R$ (Supplementary Figure 4). The exact perimeter of the resulting texture's edge (\ie the outside perimeter of the overlapping disks), $P\left(\frac{r}{R}\right)$, reads
\begin{align}
\dfrac{P\left(\frac{r}{R}\right)}{2\pi R} = 1 + \frac{r}{R} - \frac{1}{\pi}\arccos{\left(1-\frac{r^2}{2 R^2}\right)}- \frac{1}{\pi}\frac{r}{R}\arcsin{\left(\sqrt{1-\frac{r^2}{4 R^2}}\right)}.
\end{align}
Alongside geometry 1, defined by a single circular edge of radius $\Rsk$ (\ie the original skyrmion), the other two geometries presented in the main text are defined by disk radii $r = R$ (geometry 2) and $r = R/2$ (geometry 3). For comparisons, we set $R\equiv\Rsk$.

In the case of non-circular edge shape, we straightforwardly generalize the edge state counting . Note that our geometries 2 and 3 are formed by adding an outward bulge to the circular edge of geometry 1. Therefore for simplicity we define as an edge state any state that satisfies $R_{\rm max} \geq R_{\rm sk} - l$. This criterion is less precise and may slightly overestimate the number of edge states. For simplicity we examine only the eigenstates whose energy is below $\Delta_{\rm eff}/4$, recalling that the highest edge state energy is expected to be $\varepsilon^*\approx \Delta_{\rm eff}/10$. In all studied cases the energy cutoff is $\approx 0.05 \Delta_0$.

In all geometries presented here (1, 2 and 3), the variation of the perimeter was achieved by changing the overall spatial scale while keeping the number of sites per spin-flip the same.

\subsection*{Elliptic geometry}
\label{app: ellip}

The elliptic geometry is defined simply by replacing the two-disk construction of Supplementary Figure~\ref{fig: app_blobdef} by a single ellipse at the center. We use the perimeter $P(a,b)$ of an ellipse of semi-major axis $a$ and semi-minor axis $b$ given as 
\begin{equation}
P(a,b) = 4 a E\left(\sqrt{1-\frac{b^2}{a^2}}\right),
\end{equation}
where $E\left(m\right)$ is the complete elliptic integral of the second kind defined as
\begin{align}
E\left(m\right) = \int_0^\frac{\pi}{2}\sqrt{1-m^2\sin^2\theta}\,d\theta = \int_0^1 \dfrac{\sqrt{1 - m^2 t^2}}{\sqrt{1 - t^2}} dt.
\end{align}
The selection criterion for edge states is a simple generalization of the circular case (geometry 1) since the location of points on the edge of the ellipse is simply defined in cartesian coordinates. The perimeter is increased by changing the aspect ratio of the ellipse while keeping its surface area constant, starting from $p = 16$, $q = 2$ and $\lambda/a = 12$. The results and linear fits for different edge state selection criteria $l/a = 3,\, 4, \, 5$ ($l/\lambda = 0.25, \,0.33, \,0.42$) are shown in Supplementary Figure~\ref{fig: ellipse_counting}. 

The fit results are:
\begin{itemize}
\item[\tiny $\bullet$] $l/\lambda = 0.25$: $0.31(36) \pm 0.05(35) \, a^{-1}$
\item[\tiny $\bullet$] $l/\lambda = 0.33$: $0.30(95) \pm 0.13(74) \, a^{-1}$
\item[\tiny $\bullet$] $l/\lambda = 0.42$: $0.56(93) \pm 0.21(87) \, a^{-1}$.
\end{itemize}
The data is clearly more noisy than in the case of other geometries, and the fit slopes deviate from the results of the main text (even within error bars, see Fig. 3 in the main text). We ascribe the discrepancy to the varying curvature of the texture's edge that causes hybridization between the wavefunctions on the edge. Note that in geometries 2 and 3 the edge has a non-constant curvature only at isolated points (where two circles meet).

\cleardoublepage
\newpage

\section{Stability to disorder in the 2D tight-binding model}
\label{app: 2D_disorder}

The robustness of the chiral Majorana edge mode is further confirmed by an analysis of the effect of scalar disorder in the 2D tight-binding model. The disorder is a random, spatially uncorrelated variation of the on-site energy applied throughout the system: $\mu \to \mu + \delta\mu$ where $\delta\mu$ is distributed according to a normal law of mean $0$ and standard deviation $\sigma_\mu$. 

In Supplementary Figure~\ref{fig: app_disorder} we show the results for varying disorder strengths $\sigma_\mu/\Delta_0 = 0.2, \, 0.5, \, 0.7$ in the case of the $p=9$ and $\lambda/a=16$ skyrmion.

An example of a clean system's density of states is shown in Supplementary Figure~\ref{fig: counting_dos}. While for the particular clean system whose parameters exactly correspond to the disordered systems here, there are 90 edge states in the energy range $\pm 0.05 \, \Delta_0$.
The density of states of the disordered systems, including the count of edge states in the  energy range $\pm 0.05 \, \Delta_0$, namely, 90, 106, and 98 states

for Supplementary Figure~\ref{fig: app_disorder}a, b and c, respectively, confirm that the edge mode is stable to relatively high disorder strengths.

\newpage

\section{Delocalization of the edge states in a magnetic background}
\label{app: Deloc}

Yang et al (Supplementary Reference \onlinecite{Loss_Majorana_skyrmion}) predict that if a constant magnetic background is added outside the skyrmion then the edge states delocalize from the edge into this background. The reason is simply that the superconductor is gapless in the background region. This is readily verified numerically in the radial tight-binding model of the skyrmion, as shown in Supplementary Figure~\ref{fig: FSS_deloc} where we plot the spectral weight of the peak of the edge state as the size of the background region is increased. Defining the wavefunction as $\left(u_\up(r), u_\down(r), v_\down(r), v_\up(r)\right)$, we compute the spectral weight $S_{\rm peak}$ of the state's peak through the $u_\up$ component, as
\begin{align}
S_{\rm peak} \approx \dfrac{1}{W_{\rm peak}} \sum_{j \in {\rm peak}} \left\vert u_\up(j) \right\vert^2
\end{align}
where $W_{\rm peak}$ is the estimated width (\ie length in radial direction) of the peak of the edge state. We plot the results for $S_{\rm peak}$ as a function of the background region size (\ie length in radial direction) in Supplementary Figure~\ref{fig: FSS_deloc}.

\newpage

\section{Magnetic bubbles}
\label{app: Bubbles}

For our model the most important spatial feature of bubbles is that they are essentially annulus-shaped domains of uniform polarization separated by ring-shaped domain walls Supplementary Reference \onlinecite{Kiselev:2011, Tchernyshyov:2012, Buttner:2018, BernandMantel:2018}, see Supplementary Figure~\ref{fig: Sk_v_bubble_numprofiles}a. We therefore contrast such a straightforward model of a bubble with our model of the skyrmion in Supplementary Figure~\ref{fig: Sk_v_bubble_numprofiles}b,c. In Supplementary Figure~\ref{fig: Sk_v_bubble_numprofiles}b the skyrmion shows smooth oscillation between out-of-plane (blue line is $z$-component) and in-plane magnetization (orange line). In contrast, the bubble, Supplementary Figure~\ref{fig: Sk_v_bubble_numprofiles}c, has magnetization roughly out-of-plane and constant on concentric annuli (blue line is $z$-component), while at the ring-shaped domain walls between these annuli there is in-plane winding of the magnetization (orange line is the in-plane component).

For the sake of completeness, we remind that the general texture profile is $\{n_x,n_y,n_z\}\equiv\left\{\sin(f(r))\cos(q\theta),\sin(f(r))\sin(q\theta),\cos(f(r))\right\}$, with $r$ the radial coordinate and $\theta$ the polar angle, and the explicit profile for the bubble we choose as:
\begin{equation}
f\left(r\right) = \pi \sum_i g\left(r - r_0 - r_i, w \right)
\label{eq: bubbleprof}
\end{equation}
where $g\left(x, w \right) = \left(1 + \exp\left(-x/w\right)\right)^{-1}$ is an inverted Fermi function ($g(x, w) = n_F(-x, w)$), the $r_0 = 4 w$ is a global offset, and $r_i = i\left(L - p w/2\right)/p$ is the position of the $i^{\rm th}$ domain wall, while $w$ is the radial width of the domain-wall.

We show in Supplementary Figure~\ref{fig: res_w_5_p_10} the result of our radial tight-binding computations with the bubble profile of Supplementary Figure~\ref{fig: Sk_v_bubble_numprofiles}c. The momentum-resolved excitation spectrum, Supplementary Figure~\ref{fig: res_w_5_p_10}a, shows that the nearly-flat band and the effective gap survive; the wavefunction amplitude of the zero-energy mode (its partner is not shown) at angular momentum zero shows that the Majorana wavefunctions at the center and at the edge remain localized, although their detailed spatial profile has changed significantly.

In conclusion, if the bubble’s uniform magnetization domains are not too large in real space (the radial width of the domains in above calculation is 100 sites, corresponding to a lengthscale of nm to 10 nm), our results are robust. If the domains increase, one expects to reach a regime where the ferromagnetic nature of the domains dominates, and the effective gap closes. In magnetic bubble material, the size $w$ of a domain wall is estimated as $w = \sqrt{A/K}$ where $A$ is the micromagnetic exchange constant and $K$ the (effective) magnetocrystalline anisotropy Supplementary Reference \onlinecite{BernandMantel:2018}. With the typical values $A \sim {\rm pJ}.{\rm m}^{-2}$ and $K \sim {\rm MJ}.{\rm m}^{-3}$ Supplementary Reference \onlinecite{BernandMantel:2018} and Supplementary Reference \onlinecite{Wiesendanger_target_exp}, we get $w \sim 1 - 10 \,{\rm nm}$ which is consistent with our previous estimations for the radial winding number $p\lesssim 10$ and the skyrmion radius $R_{\rm sk}\sim 10 - 100\, {\rm nm}$.

\newpage

%\bibliography{SK_arxiv_v2}

\end{document}